# Structural, Vibrational, and Electronic Behavior of Two GaGeTe Polytypes under Compression.

E. Bandiello[1,*], S. Gallego-Parra[1,*], A. Liang,[2] J.A. Sans[1], V. Cuenca-Gotor[1], E. Lora da Silva[3], R. Vilaplana[4], P. Rodríguez-Hernández[5], A. Muñoz[5], D. Diaz-Anichtchenko[2], C. Popescu[6], F.G. Alabarse[7], C. Rudamas[8], C. Drasar[9], A. Segura[2], D. Errandonea[2], and F.J. Manjón[1]

[1] *Instituto de Diseño para la Fabricación y Producción Automatizada, MALTA Consolider Team, Universitat Politècnica de València, 46022 Valencia, Spain*

[2] *Departamento de Física Aplicada-ICMUV, MALTA Consolider Team, Universitat de València, 46100 Burjassot, Spain*

[3] *IFIMUP, Institute of Physics for Advanced Materials, Nanotechnology and Photonics, Department of Physics and Astronomy, Faculty of Sciences, University of Porto, Rua do Campo Alegre, 687, 4169-007 Porto, Portugal*

[4] *Centro de Tecnologías Físicas, MALTA Consolider Team, Universitat Politècnica de València, 46022 València, Spain*

[5] *Departamento de Física, Instituto de Materiales y Nanotecnología, MALTA Consolider Team, Universidad de La Laguna, La Laguna, 38205 Tenerife, Spain*

[6] *CELLS-ALBA Synchrotron Light Facility, MALTA Consolider Team, 08290 Cerdanyola del Vallès, Barcelona, Spain*

[7] *Elettra Sincrotrone Trieste, S.S. 14 - km 163,5 in AREA Science Park, 34149 Basovizza, Trieste, Italy*

[8] *Escuela de Física, Facultad de Ciencias Naturales y Matemática, Universidad de El Salvador, San Salvador, El Salvador*

[9] *Faculty of Chemical Technology, University of Pardubice, Pardubice 532 10, Czech Republic*

## Abstract

GaGeTe is a layered topological semimetal that has been recently found to exist in at least two different polytypes, α-GaGeTe ($R\bar{3}m$) and β-GaGeTe ($P6_3mc$). Here we report a joint experimental and theoretical study of the structural, vibrational, and electronic properties of these two polytypes at high pressure. Both polytypes show anisotropic compressibility and two phase transitions, above 7 and 15 GPa, respectively, as confirmed by XRD and Raman spectroscopy measurements. Although the nature of the high-pressure phases is not confirmed, comparison with other chalcogenides and total-energy calculations allow us to propose possible high-pressure phases for both polytypes with an increase in coordination for Ga and Ge atoms from 4 to 6. In particular, the simplification of the X-ray pattern for both polytypes above 15 GPa suggests a transition to a structure of relatively higher symmetry than the original one. This result is consistent with the rocksalt-like high-pressure phases observed in parent III-VI semiconductors, such as GaTe, GaSe, and InSe. Pressure-induced amorphization is observed upon pressure release. The electronic band structures of α-GaGeTe and β-GaGeTe and their pressure dependence also show similarities to III-VI semiconductors, thus suggesting that the germanene-like sublayer induces a semimetallic character in both GaGeTe polytypes. Above 3 GPa, both polytypes lose their topological features, due to the opening of the direct band gap, while the reduction of the interlayer space increases the thermal conductivity at high pressure.

# Introduction

The pursuit of the experimental confirmation of theoretically-predicted particles such as the Weyl fermions has given great impulse to the research on topological semimetals (TSMs), especially with regard to their superconductivity features [1,2]. As a part of this effort, transition metal dichalcogenides have spanned particular attention by virtue of their optical and electronic properties, which make them promising compounds for the fabrication, among others, of water splitting devices [3] and optical detectors [4], as well as in spintronics [5] and in highly efficient thermoelectrics [6].

GaGeTe is a layered compound with a very low band gap and is one of the most intriguing non-transition metal-based TSMs [7], as reflected by the scarce studies that have been performed on it. An interesting feature of the crystalline structure of GaGeTe is the presence of a germanene-like sublayer (a corrugated monolayer structure of germanium, similar to graphene), sandwiched between two GaTe sublayers of InSe or GaSe-type [8]. The germanene sublayer has a high electrical carrier mobility, for which it is deemed to be competitive with graphene for 2D nano- and optoelectronic applications [9,10]. Additionally, GaGeTe has been found to be a stable compound in air, water, and NaOH at room conditions (RC), which is clearly favorable from the point of view of applications [11], and has been recently predicted to be dynamically and thermodynamically stable at high temperature [9]. Finally, recent studies have also confirmed the potential of GaGeTe for the fabrication of 2D field effect transistors (FETs) and photodetectors, with a high ON/OFF ratio (up to $10^3$), higher than that of many other FETs based on 2D materials [12,13].

The first synthesis of GaGeTe was reported by Kra *et al*. in 1977 [14], while structural and vibrational features at room pressure (RP) were reported a few years later [11,15]. Subsequently, most studies on GaGeTe have been aimed at discovering its potential applications on the basis of its thermoelectric, electronic, and magnetic properties [16–18]. On the other hand, until recently scarce attention has been devoted to the in-depth determination of the structural peculiarities of GaGeTe and to the possible existence of unreported polytypes. In this respect, an exhaustive study on these subjects at RC has been recently published, in which the existence of at least two polytypes, α- and β-GaGeTe, has been put in evidence [19]. One of these polytypes is centrosymmetric (α-GaGeTe, space group S.G. 166, $R\bar{3}m$) and the other is noncentrosymmetric (β-GaGeTe, S.G. 186, $P6_3mc$). These polytypes are composed of the same monolayers but stacked in a different way along the hexagonal *c* axis. Interestingly, the two polytypes have been found to be energetically competitive at RC, so they often appear simultaneously in as-prepared samples. Additionally, calculations suggest the possible existence of at least a third polytype, γ-GaGeTe (S.G. 164, $P\bar{3}m1$), although it has not yet been experimentally detected [19]. The most notable features are that both polytypes are TSMs at

RC and exhibit very small thermal lattice conductivities, as confirmed by recent studies [7,19].

Among the unanswered questions about GaGeTe, one of the most interesting ones is how the structural, vibrational, and electronic properties of this compound are influenced by thermodynamic factors, for instance by pressure. In this work, we fill this gap using a combination of powder synchrotron X-ray diffraction (XRD) and Raman spectroscopy (RS) measurements, combined with *ab initio* calculations on bulk GaGeTe at high pressure (HP). Here we show that GaGeTe exhibits remarkable features, such as anisotropic compressibility and two phase transitions in the range from 0 to 18 GPa, one above 7 GPa and another above 15 GPa. We report the equation of state (EOS) and the evolution of the unit-cell parameters of the α and β phases, as well as the characterization and evolution of the zone-center vibrational modes at HP. In addition, we report the pressure dependence of the electronic band structure and the evolution of the topological and thermal features of GaGeTe upon compression thanks to *ab initio* calculations. Regarding its structural properties, GaGeTe undergoes a transition to a high-symmetry phase above 15 GPa, in agreement with the transition to a rocksalt-like structure of parent chalcogenides InSe, GaSe, and GaTe at HP. For the phase transition above 7 GPa, we propose new structures, named α'- and β'-GaGeTe, with the same space groups as α- and β-GaGeTe but with a smaller unit cell and different atomic positions. In these two new structures the coordination of the Ga and Ge atoms increases from 4 to 6. The new structures are not present in other compounds to our knowledge and are related to the tetradymite structure of the well-known $V_2$-$VI_3$ compounds, such as $Bi_2Te_3$, that are topological insulators.

## Experimental details

Single crystal samples of GaGeTe were grown using a modified Bridgman method [17] and were characterized at RP by XRD and unpolarized RS measurements, both under resonant and nonresonant conditions, thus confirming the existence of two polytypes [19]. For HP experiments, samples were loaded in a membrane-type diamond anvil cell (DAC) with diamond culets of 400 μm in diameter. A 4:1 methanol-ethanol mixture was used as the pressure transmitting medium (PTM).

HP-RS measurements were performed at room temperature in single crystal samples of GaGeTe with the same setup used for RS experiments at RP: 532 nm solid state for resonant conditions and a 633 nm He-Ne laser for nonresonant conditions, respectively. Raman spectra were collected with a Horiba Jobin Yvon LabRAM HR UV spectrometer. Edge (532 nm) or Notch (633 nm) filters have been used to cut off the laser line and a thermoelectrically cooled multichannel CCD has been used as the detector. In all measurements, a spectral resolution better than 2 $cm^{-1}$ was obtained. Phonon signals were analyzed by fitting the Raman peaks with a pseudo-Voigt profile. Ruby chips were evenly distributed in the pressure chamber and used as the pressure gauge *via* photoluminescence measurements [20].

Three different angle-dispersive HP-XRD measurements were performed at room temperature in synchrotron facilities for powder samples of GaGeTe grinded from the original single crystal samples. Experiments 1 and 2 (Alba-1 and Alba-2, in the following) were performed with monochromatic X-rays (λ=0.4246 Å and 0.4642 Å, for Alba-1 and Alba-2, respectively) at the MSPD beamline of the Alba synchrotron (Spain) [21], using Cu (ruby fluorescence) as the pressure gauge for the Alba-1 (Alba-2) experiment. The third XRD experiment (Elettra, in the following) was conducted at the Xpress beamline of the Elettra Sincrotrone Trieste (Italy), with monochromatic X-rays (λ=0.49585 Å), using an 80 μm diameter beam size and with a PILATUS3 &M (DECTRIS) detector. The ruby fluorescence was used as the pressure gauge. In all cases, the integration of 2D diffraction images was performed using the DIOPTAS software [22], while the structural analysis (Pawley/Le Bail whole-pattern fittings) was carried out with the program MAUD [23]. EoSFit7c has been used to determine the EOS of our samples [24].

The reason for the multiple HP-XRD experiments is that, for all of the analyzed samples, pure α or β phases could not be easily isolated in powder samples, except for one single case (experiment Alba-1) in which a pure α specimen was found. In fact, *in situ* RS measurements at RC in all available single crystal samples showed the presence of the α or the β polytype (and sometimes, a superposition of both) in different layers of the sample, and often also in neighboring regions of the same layer, as commented in Ref. [19]. Hence, when grinding a significant amount of sample for XRD experiments, both polytypes are inevitably mixed together. Since the β polytype is the minority phase [19], the cell parameters of the β phase as a function of pressure have been obtained by analyzing the XRD patterns of a mixture of phases, once the cell parameters of the α phase were obtained from the only pure α phase sample we could obtain (experiment Alba-1). The reason behind the coexistence of these two polytypes in the same sample is unknown, but it is related to the fact that both polytypes are energetically competitive at RP [19]. It is thus reasonable to argue that slightly inhomogeneous conditions during the synthesis of the samples (for instance, small temperature, pressure or stoichiometry gradients) may be the reason behind the coexistence of both polytypes in almost all the samples we were able to analyze.

Finally, it must be stressed that the simultaneous presence of the α- and β-GaGeTe polytypes in all but one of our samples thwarted our efforts to confirm the proposed HP phases from HP-XRD measurements, especially for the minority β polytype, due to the superposition of many Bragg reflections, along with peak broadening and merging. Therefore, most of the information regarding the HP phases of the β polytype has been obtained from RS measurements that have been performed in single crystal samples showing only the Raman modes of one of the two polytypes, as it was previously done at RP [19].

## *Ab initio* calculation details

*Ab initio* total-energy calculations at 0 K for both phases of GaGeTe were performed within the framework of density functional theory (DFT) with the Vienna Ab-initio Simulation Package (VASP) [25], using the projector augmented waves (PAW) scheme [26,27]. The valence electron configurations adopted for Ga, Ge, and Te atoms are $3d^{10}4s^24p^1$, $3d^{10}4s^24p^2$, and $5s^25p^4$, respectively. In this work, the generalized gradient approximation (GGA) with the Perdew-Burke-Ernzerhof parametrization for solids (PBEsol), as well as the Perdew-Burke-Ernzerhof

(PBE) parametrization, including dispersion corrections from Grimme (D3) to better take into account van der Waals (vdW) interactions [28], were used for the exchange and correlation energy [29,30]. We must stress that we will discuss the values corresponding to the PBEsol functional in most of this work because PBE+D3 calculations do not show any meaningful improvement in the description of the properties of GaGeTe. A dense Monkhorst-Pack grid [31] of special *k*-points (6×6×6) along the Brillouin zone (BZ) and a plane-wave basis set with an energy cutoff of 540 eV were used. All degrees of freedom, including lattice constants and atomic parameters, were fully relaxed with self-consistent convergence criteria of 0.01 eV/Å and $10^{-6}$ eV for the atomic forces and the total energy, respectively.

Lattice-dynamical properties were obtained at the Γ-point of the BZ using the direct-force constant approach in which atomic forces were computed within the PBEsol prescription [32]. This method involves the construction of a dynamical matrix at the Γ-point of the BZ. Separate calculations of the atomic forces are needed and performed by small independent displacements of atoms from the equilibrium configuration within the primitive cell, whose number depends on the crystal symmetry. Highly converged results on forces are required for the calculation of the dynamical matrix [33]. The subsequent diagonalization of the dynamical matrix provides the frequencies of the normal modes. Moreover, these calculations allow the identification of the symmetry and the eigenvectors of the vibrational modes in each structure at the Γ-point. To obtain the phonon dispersion curves along high-symmetry directions of the BZ and the one-phonon density of states, we performed similar calculations using appropriate supercells, which allow the phonon dispersion at *k*-points to be obtained commensurate with the supercell size [33]. The J-ICE software was used to plot the atomic vibrations of GaGeTe using the OUTCAR file resulting from the VASP calculations [34].

Finally, the electronic band structures of both polytypes at different pressures along high-symmetry directions were computed within the PBEsol prescription including spin-orbit coupling (SOC). With this information, the topological properties of both polytypes were calculated at different pressures, as proposed by Fu and Kane [35], with the same approach we used in a previous work [19].

## Results and Discussion

1. **Structural properties**

α-GaGeTe crystallizes in a trigonal layered structure (S.G. $R\bar{3}m$, No. 166, Z=6), with lattice parameters, in a hexagonal setting, *a* = 4.048 Å and *c* = 34.734 Å (*V*=492.91 Å$^3$) [19]. The hexagonal unit cell consists of three monolayers piled up along the c axis. Each monolayer consists of six atomic planes perpendicular to the *c*-axis, with a sequence Te-Ga-Ge-Ge-Ga-Te (see **Figure 1a**). The intralayer forces between atoms are of covalent type, while the forces between the layers are of vdW type [36]. In this layered structure, Ga is tetrahedrally coordinated with 3 Te atoms and 1 Ge atom, Ge is tetrahedrally coordinated with 3 Ge and 1 Ga atom, and Te atoms are threefold coordinated to Ga atoms. On the other hand, β-GaGeTe crystallizes in a hexagonal layered structure (S.G. $P6_3mc$, No. 186, Z=6) with *a* = 4.0379 Å and *c* = 22.1856 Å (*V*=313.27 Å$^3$) [19]. The structure of β-GaGeTe is similar to that of α-

GaGeTe, since both polytypes are formed by the same monolayers, but the hexagonal unit cell of the β polytype consists of two monolayers piled up along the *c* axis (**Figure 1b**).

The atomic arrangement is different between the two polytypes, as shown in ref. [19], since in the β polytype one layer is the specular image of its neighboring layers, unlike in the α polytype, where all layers are equally oriented. Consequently, there is only one atom of each kind in the primitive unit cell of α-GaGeTe and all atoms are located at the 6*c* Wyckoff positions, with *x/a=y/b*=0. On the other hand, there are two atoms of each kind in the primitive unit cell of β-GaGeTe, located at the 2*a* and 2*b* positions. These differences have implications on the axial compressibility of the polytypes, as we will discuss in the following. Contrary to what happens for the α polytype, however, in the β structure two different types of Ga and Ge coordination tetrahedra exist, with slightly different bond lengths and volumes. These will be referenced here as the Ga1, Ga2 and Ge1, Ge2 polyhedra, respectively, according to the notation already used in ref. [19].

First, we will comment on the results of the Alba-1 experiment, which was the only one in which pure α-GaGeTe could be analyzed (i.e., without the presence of the β polytype). The XRD pattern of the pure α-GaGeTe powder sample inside the DAC (see **Figure S1** in the Electronic Supplementary Information (ESI)) shows a strong preferred orientation (as well as in all XRD patterns of the three experiments), as expected for layered compounds [37]. Therefore, only Le Bail refinements have been performed on all the XRD patterns. In particular, the XRD pattern at 1.2 GPa can be entirely fitted using the S.G. $R\bar{3}m$ with parameters $a$ =4.0204(1.4) Å and $c$ =34.214(5) Å (see **Figure S1**). These values are in agreement with the values given previously at RP [19] (see **Table S1** in ESI).

**Figure 2** shows the XRD patterns from the Alba-1 experiments plotted at selected pressures up to 13.6 GPa. The XRD patterns barely change with increasing pressure, except for the peak shifting towards higher 2θ values due to the pressure-induced shrinking of the unit cell. However, changes occur in the XRD patterns above 5.3 GPa as evidenced by the decrease in intensity of the peak group around 7° and the merging of the peaks at approximately 15°. We believe that these changes mark the onset of a phase transition to a new structure.

In order to identify the new HP phase of α-GaGeTe, we have considered previous results from analogue chalcogenides, such as InSe, GaSe, GaTe, and $Ga_2Se_3$, whose layered structures undergo a phase transition to the rocksalt (NaCl-type) structure at HP [38–40]. The transition from a fourfold-coordinated structure, such as GaGeTe, to a sixfold-coordinated structure, such as rocksalt, is therefore expected in this material. However, XRD patterns above 5.3 GPa do not correspond to a rocksalt structure and they can still be fitted in S.G. $R\bar{3}m$. The lack of observation of new peaks and the expected tendency towards sixfold coordination suggests that a possible HP structure above 5 GPa with increased atomic coordination of the Ga atom may have the same S.G. as the original phase. As a matter of fact, our theoretical calculations demonstrate the possible occurrence of a HP phase of the α polytype in a structure within the same space group than the original phase, S.G. $R\bar{3}m$, but with sixfold coordination for Ga and Ge. This HP phase has been named $\alpha'$-GaGeTe and, as shown in **Figure** S2 in ESI, is energetically competitive with the original $\alpha$ phase at a pressure above 8.7 GPa. The proposed phase is a layered rocksalt structure not found previously in any compound and it bears some similarity with the layered tetradymite structure present in the topological insulator $Bi_2Te_3$ at RP. Details of the structural parameters of the proposed α'-GaGeTe are provided in **Table S2** in ESI and a scheme of the structure is shown in **Figure S3a** in ESI. For comparison purposes, the $Bi_2Te_3$ structure is shown in **Figure S3c** in ESI. As observed in **Figure S3**, the proposed α' phase is similar to that of $Bi_2Te_3$, but with the central Te sublayer substituted by the germanene sublayer. The proposed α'-GaGeTe would occur by breaking the Ga-Ge bond. This will produce the collapse the GaTe sublayer over the germanene sublayer so that the Ga atom occupies an octahedral site in order to form three new Ga-Ge bonds. In this way, both Ga and Ge atoms change from fourfold to sixfold coordination. Similarly, an analogous $\beta'$ structure considered as the HP phase of $\beta$-GaGeTe and within the same space group of the low-pressure phase, S.G. $P6_3mc$, but with sixfold coordination for Ga and Ge (see **Table S3** and **Figure S3b** in ESI), appears to be energetically competitive with the low-pressure $\beta$ phases at a pressure above 9.2 GPa (see **Figure S2** in ESI). The $\alpha'$ (resp., $\beta'$) structure cannot thus be excluded as the possible HP phases of $\alpha$-GaGeTe (resp., $\beta$-GaGeTe) at a pressure above 8.7 GPa (resp., 9.2 GPa).

In Alba-2 and Elettra XRD experiments, a mixture of both α and β polytypes was initially present in the examined samples as shown in XRD patterns at selected pressures in **Figures 4 and 5**, respectively. Le Bail fits of the XRD patterns of the Alba-2 and Elettra experiments at nearly-ambient pressure are shown in **Figures S4a and S4b**, respectively, where a peak attributable to the ruby used as the pressure gauge is also visible at low 2θ. The cell parameters obtained from these experiments for both polytypes of GaGeTe (**Table S1**) are in agreement among them and with those reported previously [19]. As these experiments include contributions from both polytypes, it is not straightforward to attribute the changes observed in the patterns to a specific polytype. For the Alba-2 experiment, changes in the XRD patterns, in particular the decrease of the intensity of some of the peaks, are visible at a pressure above 7.8 GPa, as evidenced by the arrows in **Figure 4**. These changes are indicative of the onset of a phase transition, coherently to what was observed in the Alba-1 experiment. It has to be mentioned that for the Alba-2 experiment the XRD patterns above 10.6 GPa suffer from a huge broadening, maybe due to bridging in addition to the expected loss of hydrostaticity of the PTM above this pressure (**Figure 3**) [41]. Due to this, Le Bail refinement of Alba-2 patterns above 10.6 are unreliable and have not been taken into account in our analysis. However, these patterns and those relative to the pressure release are equally shown in **Figure 3** for completeness. Qualitatively speaking, it can be noted that only a single broad peak at 2θ ≈ 16° is present in the XRD patterns around 16.4 GPa, which may indicate a second phase transition with a further increase in the crystalline symmetry (**Figure 3**). This peak shifts towards higher values of 2θ upon further pressure increase. Upon decompression, at 9.2 GPa, a less intense peak appears besides the main peak, at a lower 2θ, and it survives down to 3.6 GPa. This peak may be related to a partial recovery of the previous HP phase. This intermediate state is unstable since pressure-induced amorphization (PIA) takes place below 1.3 GPa (**Figure 3**). The second transition is thus not reversible. Similar results can be observed in the Elettra experiment, (**Figure 4a-c**). The XRD patterns at low pressure are all similar up to 7.5 GPa when, again, the intensity of some peaks starts to decrease (**Figure 4a**). Starting from a pressure of 14 GPa new changes can be observed in the XRD patterns, such as the appearance of new peaks and the complete quenching of some Bragg reflections at 15 GPa (**Figure 4b**), thus suggesting, again, a further symmetry increase. The patterns at 15 GPa and above are broadened by the non-hydrostaticity of the PTM. This fact, together with the mixture of α and β polytypes, makes it extremely difficult to determine the actual structure of the HP phase above 15 GPa. Nonetheless, the reduced number of peaks at 15 GPa and above (**Figure 5**) is symptomatic of a phase with a highly symmetrical structure, maybe tetragonal or even orthorhombic. This fact suggests that that GaGeTe may follow a sequence of pressure-induced phase transitions similar to that observed for GaSe, $Ga_2Se_3$, GaTe, and InSe, that exhibit a fourfold to sixfold increase in atomic coordination due to the appearance of the rocksalt-like structure above 10 GPa [38–40]. In any case, more experimental work, ideally under nearly-hydrostatic conditions and with a single phase, is necessary to confirm the HP phases of GaGeTe polytypes above 15 GPa.

The intensity of the new peaks observed above 15 GPa increases up to around 18.5 GPa, the maximum pressure reached in this experiment (**Figure 4b**). The latest structure is retained upon decompression up to a pressure of 3.9 GPa, when PIA finally occurs at a pressure close to that observed in the Alba-2 experiments. As a last remark, the groups of peaks at low 2θ in **Figures 2, 3, and 4a** (≈7°, 8° and 8°, respectively) apparently survive to the onset of the first phase transition, although their intensity is strongly attenuated. This suggests that the low-pressure and the first HP phase may coexist in the bulk in a given pressure range (apparently, up to 14 GPa, **Figure 4b**). This result is consistent with a first-order phase transition, as expected for the proposed α-to-α' phase transition.

Following our Le Bail fits, we were able to study in detail the compressibility of both polytypes of GaGeTe at 0 GPa. **Figure 6** shows the pressure dependence of the structural parameters of both polytypes according to the three performed XRD experiments. Firstly, we will discuss the axial compressibility of the structures, defined for each axis, namely $x$, as $\chi_x = -\frac{1}{x_0}\frac{\partial x}{\partial P}$ (with $x_o$ being the value of $x$ at RP). In **Figure 6a** and **6b**, we show the dependence of the lattice parameters of α-GaGeTe and β-GaGeTe. Those figures evidence that the quality and dispersion of the parameters obtained by LeBail fits worsen rapidly at pressures above 8 GPa, due to peak superposition and broadening. Thus, although for completeness all of the obtained data are shown in **Figures 6a** to **6d**, only the data up to 8 GPa have been used for the determination of axial and bulk compressibility of both polytypes of GaGeTe. The plots show that the compressibility of both polytypes is strongly anisotropic, as expected for layered vdW compounds [38–40,42,43]. The experimental and calculated values of the axial compressibility of α- and β-GaGeTe, (**Table 1**), are in very good agreement. However, it can be noticed that the experimental values of $\chi_c$ for β-GaGeTe deviate significantly from the theoretical calculations. This may be due to a slight theoretical overestimation of the initial value of the $c$ parameter of the β polytype (**Table S1** and **Figure 6b**). Additionally, the calculated $\chi_c$ is higher for β-GaGeTe than for α-GaGeTe, despite the lower initial volume of the former (mainly due to the smaller initial value of the $c$ parameter). The anisotropy and zero pressure compressibility of the lattice parameters in both polytypes of GaGeTe is comparable to those of parent layered compounds InSe, GaSe, GaTe, $Ga_2Se_3$ and $In_2Se_3$ [38–40,42,43].

The bulk compressibility of both polytypes of GaGeTe are shown in **Figure 6c**. Experimental and calculated data of unit cell volumes are in good agreement, although the unit cell volume at RP, $V_0$, of β-GaGeTe is slightly overestimated by calculations, as a result of the overestimation of the $c$ parameter, as previously mentioned. All the experimental and theoretical $P$-$V$ data have been fitted using a 2$^{nd}$ order Birch-Murnaghan EOS [44], to facilitate the comparison of the bulk modulus data, $B_0$. The EOS fit of the experimental data gives $B_0$= 41.6(7) and 48.5(8) GPa for α- and β-GaGeTe, respectively, thus the β polytype is actually the least compressible of the two, as expected due to its smaller $V_0$ and the inverse relation between $V_0$ and $B_0$. Notably, theoretical calculations give similar $B_0$ values for both polytypes, around 44.5 GPa (see **Table 2**). At this point, we consider that the bulk modulus of the β polytype is underestimated, probably due to the aforementioned overestimation of the initial volume $V_0$, resulting from the overestimation of the $c$ lattice parameter. Although a direct comparison of $B_0$ values of GaGeTe polytypes with isostructural compounds with analogue stoichiometry is not possible, we can mention that they are comparable to those of layered parent compounds, such as InSe, GaSe, GaTe, φ-$Ga_2S_3$, and α-$In_2Se_3$ [38–40,42,43]. Summarizing, α- and β-GaGeTe are vdW layered compounds with very pronounced anisotropic compressibility, much higher along the $c$ axis than along the $a$ axis, and are soft materials ($B_0$ < 50 GPa) as the analogue layered compounds.

Due to the good agreement between the pressure dependence of experimental and theoretical lattice parameters and unit-cell volume and to the impossibility to perform Rietveld refinement on HP-XRD data, we have used theoretical calculations to study the pressure dependence of the atomic free parameters (**Figure S5** in ESI) and of interatomic distances (**Figures S6a and S7** in ESI). With regard to the atomic free parameters, the theoretical values at RP show a good agreement with the experimental data. With regard to the pressure dependence of the theoretical interatomic distances, it can be observed that intralayer distances (Ga-Te, Ga-Ge, Ge-Ge) in α-GaGeTe at RP are well reproduced by our calculations. Moreover, the Ga-Te, Ge-Ge, and Ga-Ge bond distances in α-GaGeTe at RP are similar to the Ga-Te, Ge-Ge, and Ga-As bond distances in crystalline GaTe, Ge, and GaAs [45–47].

Upon compression at moderate pressure (below 3 GPa), the decrease in the unit-cell volume in both polytypes is mainly driven by the reduction of the $c$ lattice parameter. By comparing the theoretical evolution under compression of the different atomic parameters along the $c$ axis, it is evident that the interlayer separation between two neighbor monolayers (i.e., the projection along the $c$ axis of the distance between the two nearest Te atoms belonging to two different monolayers) is initially very similar between the two polytypes and evolves in a similar way under compression (**Figure S7**). The reduction in the Te-Te interlayer distance between the monolayers as pressure increases is much larger than the reduction in the monolayer thickness or the intralayer Ga-Ga thickness in this pressure range (see **Figure S7**). Therefore, the strong decrease of the $c$ lattice parameter at low pressures is dominated by the strong reduction of the Te-Te interlayer distance in the pressure range below 3 GPa.

The only substructure whose behavior significantly differs in compression between the two polytypes is the germanene-like sublayer. In fact, the thickness of this sublayer is initially higher in β-GaGeTe than in α-GaGeTe; i.e., its corrugation is larger in β-GaGeTe than in α-GaGeTe. At the same time, the thickness of the germanene-like sublayer decreases more steeply under compression for the β polytype than for the α polytype (**Figure 7**). This may be thus the reason behind the higher calculated value of $\chi_c$ for the β than for the α polytype. On the other hand, it can be observed that the calculated value of the thickness of the germanene-like sublayer reaches a minimum around 8 GPa for the α polytype and above 10 GPa for the β polytype. This minimum is correlated with the calculated minimum of the *c/a* ratio (**Figure 6d**) in both polytypes, since it occurs around 6 GPa for the α polytype and between 8 and 9 GPa for the β polytype. In both cases, this corresponds to a relative volume change of ~0.89 and ~0.86 (for α and β, respectively) compared to the volume at RP. This change in the trend of the *c/a* ratio, observed in several layered compounds and attributed to a change in the chemical bonding [48,49], could be related to the phase transition taking place in this pressure region in GaGeTe. In this context, it must be stressed that the increase of the thickness of the germanene sublayer following its minimum, occurs, for both polytypes, despite the decrease in the Ge-Ge bond distance, as a consequence of the decrease of the Ge-Ge-Ga and Ge-Ge-Ge angles; i.e. due to the increase of the corrugation of the germanene-like layer above certain pressure. Therefore, the increase of the thickness of the germanene sublayer above certain pressure is likely due to the stiffening of the interlayer vdW interaction, which in turn supports the structural instability that triggers a phase transition above 6-8 GPa.

Finally, as already mentioned, let us remind that, within the single layers, the coordination tetrahedra of Ga and Ge show a single value of Ga-Ge, Ge-Ge, and Ga-Te bond lengths in α-GaGeTe and slightly different bond lengths in β-GaGeTe (**Figure S6a**, **Table S4** in ESI). The plots in **Figure S6a to S6d**, regarding the theoretical parameters of the Ga polyhedra (bond lengths, volume, distortion index, and effective coordination) confirm the similarities between the Ga and Ge tetrahedra in both polytypes, along with their comparable compressibilities. As shown in **Table S5** and as obtained from **Figure S6b**, the values of the bulk modulus, $B_0$, for Ga tetrahedra are 68.7(4) in α-GaGeTe and 69.2(3) and 73.11(17) GPa for Ga1 and Ga2 tetrahedra in β-GaGeTe, respectively. Similarly, for the Ge tetrahedra, $B_0$ equals 60.6(3) GPa in α-GaGeTe and 62.75(18) and 62.81(18) GPa for the Ge1 and Ge2 tetrahedra in β-GaGeTe, respectively. This result means that the bulk moduli of the Ga and Ge coordination tetrahedra in both polytypes is around 25%-30% higher than the bulk moduli of the whole structure. Since the bulk moduli of Ga and Ge tetrahedra are given by the compression of intralayer units, the smaller bulk modulus of the whole structure is consistent with the $B_0$ value being mainly given by the reduction of the interlayer distances between the monolayers and not by the reduction of intralayer distances (see **Figure S7** in ESI).

Summarizing, all XRD experiments show an anisotropic compressibility and the occurrence of at least two phase transitions for α-GaGeTe, one at a pressure above 7 GPa and the second above 15 GPa. PIA occurs upon complete pressure release, thus confirming that the phase transitions are not reversible beyond 15 GPa. The mixture of polytypes in the original samples made it difficult to unambiguously determine the two HP phases. However, XRD patterns above 15 GPa exhibit features that suggest a transition to a more symmetric structure, as observed in analogous layered compounds. On this basis, we have proposed, for both polytypes and in the range from 7 to 15 GPa, phases energetically competitive (α' and β') with the original ones (α and β) but with a sixfold cation coordination for both Ga and Ge atoms instead of the original fourfold coordination. These phases still have layered arrangements and resemble the tetradymite structure. Moreover, the first HP phase transition is of first order, given the coexistence of the low and HP phases in a given pressure range.

To conclude this section, it is important to remark that additional studies are needed in order to clarify some aspect of the structural behavior of GaGeTe under compression. The first step in this respect would certainly be to consistently grow pure α- and β-GaGeTe samples, thus allowing more reliable XRD data under compression for each polytype. Equally more important would be to perform XRD experiment under compression using a hydrostatic PTM (for instance, helium). In fact, the most relevant changes in the XRD patterns i.e., the two phase transitions, occur at a pressure close and above the hydrostatic limit of the 4:1 methanol-ethanol mixture used as PTM. Consequently, the broadening of the Bragg peaks, along with the temporary coexistence of the low and HP phases (as mentioned before) complicates the determination of the actual HP phases. This is important in view of the confirmation of the high-symmetry phase expected at HP, i.e., above 15 GPa, as well as the exact determination of the intermediate structures. We hope that our results will stimulate further research aimed at determining the structural behavior and exact phase transition sequence of GaGeTe under compression.

## 2. Vibrational Properties

The RS spectrum of α- and β-GaGeTe at RC has been recently studied in detail, under resonant and nonresonant excitation conditions [19], thus solving unanswered questions remaining from an old RS work [15]. Therefore, here we will discuss the pressure dependence of the Raman-active modes of both polytypes using the assignments and notations given in ref. [19]. According to group theory, α-GaGeTe has 6 Raman-active modes in the wavenumber range below 350 cm$^{-1}$, $3E_g+3A_{1g}$, 4 IR-active modes, $2E_u+2A_{2u}$, and acoustic modes, $1E_u+1A_{2u}$ [50]. On the other hand, β-GaGeTe has 16 Raman-active modes in the same wavenumber range, $5A_1+5E_1+6E_2$, 10 IR modes, $5A_1+5E_1$, 6 silent modes, $6B_1$, and acoustic modes, $1E_1+1A_1$ [19,50]. Due to the similarity of the structure of both polytypes, the modes of β-GaGeTe can be related to those of α-GaGeTe, by folding its phonon dispersion branches along the Γ−Z direction of the BZ. In this way it can be shown that all 6 $A_1$ and 6 $B_1$ modes of β-GaGeTe come from all the 3 $A_{1g}$ and 3 $A_{2u}$ modes of α-GaGeTe and the 12 $E$-type modes of the β polytype come in 6 pairs ($E_1,E_2$) that derive from the $3E_g$ and $3E_u$ vibrational modes of the α polytype [19]. It is important to remark that, for both polytypes, modes with $E$ symmetry are doubly degenerated and related to vibrations of the atoms perpendicular to the hexagonal $c$ axis, while modes with $A$ or $B$ symmetry are related to vibrations of atoms parallel to the $c$ axis [19].

**Figures 8a and 8b** show the RS spectra of α-GaGeTe and β-GaGeTe, respectively, at selected pressures as measured with the green laser (resonant excitation conditions at RP). These RS spectra show less traces of Fermi resonance in α-GaGeTe than RS spectra measured with the red laser (nonresonant excitation conditions at RP) which are shown in **Figures S8a and S8b** in ESI [19]. It is interesting to note that RS spectra for pure α and β polytypes could be easily acquired, contrarily to what happened for powder XRD experiments. This is due to the local character of RS measurements, where only a few square microns at the surface of each sample are analyzed (GaGeTe is opaque to both red and green light). In some occasions, it was possible to observe the presence of both polytypes in the same sample, most likely at the edge of two domains; however, it was relatively easy to select one of the two polytypes by moving the laser to a different nearby location of the sample.

In the present work, the 6 Raman-active modes for α-GaGeTe could be detected experimentally at RP and followed at HP, while only 8 Raman-active modes could be identified for β-GaGeTe (see **Figures S9a and S9b** for identification of modes of both polytypes at low pressures inside the DAC when measured with the green laser). The hallmark feature of the RS spectrum of α-GaGeTe inside the DAC near RP is the $A_{1g}^1$ mode around 81 cm$^{-1}$. Instead, a prominent $A_1^1$ peak close to 66 cm$^{-1}$ is the signature of β-GaGeTe. Additionally, a low frequency $E_g^1$ mode is expected for α-GaGeTe around 40 cm$^{-1}$ at room pressure [19]. This mode could not be observed in the spectra acquired below 2.5 GPa because it is cut off by the edge filter. However, it becomes visible above 2.5 GPa, due to its positive pressure coefficient (see **Figure 8a**). In this regard, all experimentally detected Raman-active modes of both polytypes show positive zero-pressure coefficients, except for the $E_1^2$ mode of β-GaGeTe around 59 cm$^{-1}$ (see **Tables 3 and 4**). Changes in the RS spectra have been noted above 7 and 15 GPa in both polytypes in **Figures 8 and S8**. The origin of these changes will be discussed at the end of this section. Now we will focus on discussing the pressure dependence of the Raman-active modes of both polytypes below 7 GPa.

As already commented in our previous work [19], a Fermi resonance of the two phonons of the middle wavenumber region (modes $E_g^2$ and $A_{1g}^2$) is observed for RS measurements of α-GaGeTe excited under nonresonant conditions (with the red laser). This resonance effect can be observed at all pressures up to 7 GPa (see the broad bands with several maxima between 100 and 170 cm$^{-1}$ represented in **Figure S8a**). However, this effect is absent in measurements performed with the green laser at RP i.e., under resonant conditions. Notably, the Fermi resonance is observed exciting with the green laser when resonant conditions are lost at HP (see **Figure 8a** above 2.9 GPa). On the other hand, Fermi resonance is not observed in RS measurements of β-GaGeTe irrespectively of the excitation conditions. The conditions for the observation of the Fermi resonance were previously described [19].

A comparison of the experimentally detected and theoretically predicted Raman-active mode zero-pressure wavenumbers and pressure coefficients, as measured with the green laser, is shown in **Tables 3 and 4**, for α-GaGeTe and β-GaGeTe, respectively, as well as in **Figure 9**. For α-GaGeTe, most of the modes originated by the Fermi resonance are not represented, since a detailed discussion was already performed in this regard in Ref. [19]. A rather good agreement between experimental and theoretical results for the wavenumbers and pressure coefficients of the Raman-active modes at RP and at HP is observed for both polytypes. The absolute difference between theoretical and experimental wavenumbers for the different vibrational modes usually increases with their wavenumber, although most modes are in agreement within a 5% uncertainty. The largest deviation was found for the phonon of β-GaGeTe at 170 cm$^{-1}$, attributed to the $A_1^2$ mode, whose theoretical value deviates more than 10% from the experimental value. Therefore, we can confirm the symmetry assignment of the experimental modes observed in both polytypes and already reported in Ref. [19]. For completeness, we provide the theoretical pressure dependence of the IR-active modes of α-GaGeTe and silent $B_1$ modes in β-GaGeTe in **Tables S6 and S7** as well as in **Figures S10a** and **S10b**, respectively, in ESI.

The reason why not all Raman active modes of β-GaGeTe are detected could be related to the degeneracy of wavenumbers of many of the $E_1$ and $E_2$ modes, but it could also be that only $E_1$ modes have been detected because $E_2$ modes are quite weak and most modes that can be either $E_1$ or $E_2$ show a pressure dependence similar to that of the predicted $E_1$ modes. A confirmation of the difference between both polytypes is given by: i) the detection of a soft Raman-active mode in β-GaGeTe, correlated with a soft IR-active mode in α-GaGeTe; and ii) the detection of the anticrossing of the two lowest $E_1$ modes in β-GaGeTe near 9 GPa (see **Figure 9b**). According to our theoretical calculations, this $E_1$-mode anticrossing can only occur in the β polytype (calculations predict it around 7 GPa due to uncertainties in the absolute wavenumber values). All in all, the observation of the anticrossing confirms the different nature of both α and β polytypes.

With regard to the pressure dependence of vibrational modes in α-GaGeTe **(Figure 9a and S10a)**, it is evident, in first place, that the dependence of phonon wavenumbers on pressure is not linear in the range from RP to 7 GPa so in this range we have fitted them to quadratic functions (see pressure coefficients in **Table 3 and Table S6**). As already mentioned, all Raman-active modes of α-GaGeTe have positive pressure coefficients. At low frequency, the $A_{1g}^1$ mode at 79 cm$^{-1}$, related to the vibration of the layers against each other along the *c* axis (compressional mode) [19], has one of the largest pressure coefficients (5.9 cm$^{-1}$ GPa$^{-1}$), together with the largest, by far, Grüneisen parameter, γ (2.8). This is consistent with the anisotropic compressibility of α-GaGeTe and remarks that compression is dominated, at least initially, by the strong reduction in the interlayer spacing, as mentioned in the previous section. The large pressure coefficient of the $A_{1g}^1$ mode is due to the steep increase in the interaction between adjacent layers, caused by the strong reduction in the interlayer distance that occurs in layered vdW compounds, such as InSe and GaSe. In fact, similar values of the pressure coefficients and Grüneisen parameters for the compressional mode can be observed in α-GaGeTe and in InSe and GaSe [51,52]. A similar behavior is found in the layered Rashba semiconductor BiTeBr [49] and in layered topological insulators, β-As$_2$Te$_3$, Bi$_2$Se$_3$, Bi$_2$Te$_3$, and Sb$_2$Te$_3$ [53–57] despite these layered compounds do not show pure vdW interlayer interactions [56,57]. The large γ value for the $A_{1g}^1$ mode indicates the strong anharmonic contribution of interlayer forces. On the other hand, the small pressure coefficient of the shear or transversal layered mode $E_g^1$ mode at 41 cm$^{-1}$ is due to the weak transversal component of interlayer vdW forces between the neighboring layers [19]. This mode has the second largest γ value among all the Raman modes of α-GaGeTe due to the anharmonic contribution of interlayer forces.

The medium-wavenumber $E_g^2$ and $A_{1g}^2$ Raman-active modes at 179 and 206 cm$^{-1}$, respectively, have pressure coefficients between 2 and 3.5 cm$^{-1}$ GPa$^{-1}$. The former is dominated by the Ga-Ge bending mode and a Ga-Te stretching mode, while the latter is predominantly a mixture of the Ge-Ge bending mode (thus, an oscillation of the thickness of the germanene-like sublayer) and a Ga-Te bending mode [19]. For these phonons, it must be noted that a rigid interlayer shear ($E_g^2$) and compressional ($A_{1g}^2$) contribution exist, being both very weak because of the small amplitude of the out-of-phase vibration of Te atoms of neighbor layers. Finally, there are two high-wavenumber Raman-active modes. The $E_g^3$ mode of α-GaGeTe at 280 cm$^{-1}$ is mainly contributed by the in-plane Ge-Ge stretching mode, which gives $E_g^3$ the largest pressure coefficient of the α polytype, while the $A_{1g}^3$ mode (295 cm$^{-1}$) owes its relatively large pressure coefficient to the Ga-Ge symmetric stretching (Ga and Ge are aligned along the hexagonal *c* axis and are thus sensitive to the strong compression of the *c* axis at low pressures) with a smaller contribution given by the Ge-Ge bending mode [19]. Due to the similar masses of the atoms involved in the stretching components (Ga and Ge), these two latter modes have similar frequencies across the entire 7 GPa pressure range. The decrease in the thickness of the germanene-like sublayer at HP, in the range from RP up to 8 GPa (**Figure 7**), is consistent with the positive pressure coefficient of the $A_{1g}^2$ and $A_{1g}^3$ modes. Except for $E_g^1$ and $A_{1g}^1$, all other Raman modes of α-GaGeTe show comparable, relatively low Grüneisen parameters, $\gamma \lesssim 1$, indicating a moderate anharmonicity of most Raman modes.

Regarding the IR-active modes of α-GaGeTe (**Figure S8a**), it is worth mentioning the negative pressure coefficient predicted for the IR $E_u^1$ mode. Since this mode is a rigid intralayer mode between the germanene-like sublayer and the GaTe sublayers [19], its negative pressure coefficient suggests that the Ga-Ge bond loses electrical charge as pressure increases i.e., that the Ga-Ge bond decreases in strength under compression. This hypothesis allows also to explain the small pressure coefficient of the $A_{2u}^1$ mode (**Figure S8a**); a mode also related to the Ga-Ge bond distance that otherwise would be expected to have a large pressure coefficient due to the considerable decrease in the Ga-Ge bond distance at HP (**Figure S6a**). In this context, it must be stressed that low-wavenumber Raman-active, IR-active modes and BZ-edge modes with negative pressure coefficients are commonly observed in many materials with structures with a tetrahedral atomic coordination, related or derived from the cubic diamond, wurtzite, and zincblende structures, such as Si [58], ZnO [59], CdGa$_2$Se$_4$ and CdAl$_2$S$_4$ [60,61], to name a few, as well as in zircon-type ABO$_4$ compounds [62,63]. This anomalous decrease in the phonon wavenumber at HP is not caused by an increase of the cation-anion distances, but by a decrease in the electrical charge of the bond and it has been related to the instability of these structures derived from the cubic diamond lattice with respect to the octahedral coordination at HP [64]. In cubic lattices, the soft modes that show negative pressure coefficients usually occur at the BZ edge (see Ref. [64] and references therein). However, in crystalline compounds with a lower lattice symmetry or with a large number of formula units per primitive cell, the folding of the unit cell along certain low-symmetry directions into the BZ center (Γ-point), allows these BZ-edge vibrations to be located at the Γ-point i.e., to become Raman- or IR-active modes and even silent modes. The results presented above for α-GaGeTe suggest that the breaking of the Ga-Ge bond due to a shear mode could be involved in the first pressure-induced phase transition around 7 GPa. In fact, the proposed HP phase α'-GaGeTe, in which Ga and Ge are sixfold coordinated, is the most symmetric phase obtained by the shear displacement of the GaTe sublayers with respect to the germanene-like sublayer. As commented in the previous section, this phase transition requires the breaking of the Ga-Ge bond in α-GaGeTe so that Ga atoms can locate in octahedral sites of the $R\bar{3}m$ structure, in order to give rise to three new Ga-Ge bonds and the corresponding change from fourfold to sixfold coordination for both Ga and Ge atoms.

With regard to the pressure dependence of vibrational modes in β-GaGeTe **(Figure 9b and Figure S8b)**, similarly to what happens for α-GaGeTe, the dependence of phonon wavenumbers on pressure is not linear in the range from room pressure to 7 GPa, so we have fitted them to quadratic functions in this range (see **Table 4 and Table S7**). In β-GaGeTe, all predicted Raman-active modes have positive pressure coefficients, except for three modes ($E_1^2, E_2^2$ and $E_2^3$). Of these three soft modes, we were able to measure only the $E_1^2$ mode near 59 cm$^{-1}$. This mode and the $E_2^3$ mode are shear intralayer modes between the germanene-like sublayers and the GaTe sublayers [19], so they are related to the $E_u^1$ mode just discussed in α-GaGeTe. Therefore, the observations made above for the decrease in the Ga-Ge bond strength at HP and the possible first HP phase transition of α-GaGeTe can also be applied to β-GaGeTe. For this reason, here we propose that the first HP phase of β-GaGeTe above 7 GPa, β'-GaGeTe, could be analogous to the β phase but with a slightly different atomic arrangement and sixfold coordination for Ga and Ge, as anticipated in the "Structural properties" Section (see details in **Fig. S3b** and **Table S3**). A consequence of the negative pressure coefficient of $E_1^2$ is the appearance of an anticrossing with mode $E_1^1$ (34 cm$^{-1}$), as previously commented. The $E_1^1$ and $E_1^2$ modes are both shear intralayer modes with Ga-Ge bending nature in which GaTe sublayers and the germanene-like sublayer vibrate out-of-phase (in fact, in the $E_1^1$ mode the germanene sublayer is at rest). However, the $E_1^1$ mode is also partially a shear interlayer mode in which Te atoms of neighbor layers vibrate out-of-phase [19]. Therefore, this mode has a slightly positive pressure coefficient, similar to the $E_2^1$ mode in β-GaGeTe. As regards the $E_2^2$ mode, it cannot be detected experimentally and its calculated frequency is 35.6 cm$^{-1}$ at 0 GPa. This mode has a very low (theoretical) pressure coefficient of -0.36 cm$^{-1}$ GPa$^{-1}$, while its associated $E_1^1$ mode [19] has the lowest positive pressure coefficient among all modes of β-GaGeTe. It is worth mentioning that both related $E_1^1$ and $E_2^2$ modes come from the $E_g^1$ mode in α-GaGeTe (in fact, the $E_1^1$ mode has exactly the same atomic vibrational pattern as the $E_g^1$ mode), which also shows the smallest positive pressure coefficient in α-GaGeTe.

The largest pressure coefficients of the low-wavenumber modes of β-GaGeTe correspond to the $B_1^1$ and $A_1^1$ modes. The first one is the longitudinal or compressional layer mode and has the largest pressure coefficient and Grüneisen parameter. The second one is a longitudinal intralayer mode of β-GaGeTe in which the GaTe sublayer vibrates against the germanene-like sublayer and also against the GaTe sublayer of the neighbor layer (so it has a partial compressional layer character as the $B_1^1$ mode) [19]. The reasons for the large pressure coefficients and Grüneisen parameters for these two modes of β-GaGeTe are the same as for the $A_{1g}^1$ mode of α-GaGeTe since these modes in both polytypes share the same vibrational features.

All modes of β-GaGeTe in the medium-wavenumber range (100 cm$^{-1}$ < $\omega_0$ < 250 cm$^{-1}$), whether experimentally detected or just theoretically calculated, have pressure coefficients between 2.5 and 5 cm$^{-1}$ GPa$^{-1}$, just like the modes of α-GaGeTe in the same frequency range. This happens for the same reasons as in α-GaGeTe i.e., these modes correspond mainly to bending modes. Finally, all modes of β-GaGeTe in the high-wavenumber range ($\omega_0$ > 250 cm$^{-1}$), whether experimentally detected or just theoretically calculated, have high pressure coefficients, larger than 5 cm$^{-1}$ GPa$^{-1}$, exactly as the modes of α-GaGeTe in the same wavenumber range, and for the same reasons i.e., they correspond mainly to stretching modes. Among these, $E_1^5$ and $E_2^6$ can be qualitatively considered similar to $E_g^3$ mode of α-GaGeTe i.e., they are mainly contributed by vibration of the Ge atoms (Ge-Ge asymmetric stretching) perpendicularly to the $c$ axis, with minimal contributions of the Ga atoms (Ga-Ge bending), which vibrates with much lower amplitude [19]. The difference between these two modes is that atoms vibrate either in-phase ($E_2^6$) or out-of-phase ($E_1^5$) in adjacent layers. Finally, modes $A_1^4$, $A_1^5$, $B_1^5$, and $B_1^6$ mainly correspond to Ga-Ge stretching modes, in which these atoms vibrate along the $c$ axis [19]. It is therefore clear that the large pressure coefficients of the high-wavenumber modes are related to the asymmetric Ga-Ge and Ge-Ge stretching modes.

Regarding the Grüneisen parameters found for the β polytype, most of them have small positive values, γ≲1, as in the case of α-GaGeTe. The exception is the $E_2^1$ mode, expected at 17.2 cm$^{-1}$ but not experimentally detected, for which γ holds a huge value 8.4. This mode is related to movements of the layers vibrating almost rigidly in an out-of-phase fashion and it derives from acoustic modes in α-GaGeTe due to folding of the BZ edge onto the BZ center in β-GaGeTe [19]. The large γ value for this mode is thus due to the strong increase in frequency of the acoustic phonons that is caused by the strong increase in the interaction between Te atoms of adjacent layers upon compression due to the strong reduction of the interlayer space at low pressures. Contrary to what happens in α-GaGeTe, some of the modes of the β polytype have small negative γ values; in particular modes $E_1^1$, $E_2^2$ and $E_2^3$ have -1.1≲ γ≲-0.4, suggesting a moderate anharmonic character of these modes [19].

To conclude this section, we will comment on the HP phases of α-GaGeTe and β-GaGeTe. As already mentioned, changes in RS spectra occur around 7 and 15 GPa in both α and β polytypes, in correspondence with changes found in HP-XRD measurements. The easiest to explain are the changes around 15 GPa, since both polytypes show no Raman activity above this pressure. This result may be explained by a second phase transition of both polytypes to a more symmetrical phase, maybe related to a rocksalt structure, as also suggested by HP-XRD experiments for α-GaGeTe. In fact, high-symmetry phases may be Raman inactive and rocksalt-like or its variants may show metallic behavior, as in related layered chalcogenides InSe and GaSe, [65–68], what would explain the lack of Raman activity of the HP phase of GaGeTe above 15 GPa. It must be mentioned that upon decompression the Raman activity is not recovered, not even partially. This is related with the PIA observed in XRD experiments upon pressure release (**Figure 3** and **Figure 4c**).

In this context, we must also mention that a complete disappearance of Raman activity perhaps could not occur in the HP phase of GaGeTe above 15 GPa even if it is related to the rocksalt structure. It must be noted that the odd number of cations and anions in GaGeTe will likely generate considerable disorder in the HP phase above 15 GPa if a rocksalt-like structure (typical of *AX* compounds with the same number of cations and anions) is formed. It is known that disorder could lead to the observation of broad Raman bands usually related to the one-phonon density of states through disorder-induced or defect-activated Raman scattering (DARS), as it occurs in IR absorption [69–71]. However, a complete disappearance of the Raman activity even in the case of strong disorder in the HP phase above 15 GPa cannot be discarded because the possible metallic character (and perhaps high carrier mobility) of that HP phase could lead to a complete electron-phonon damping of the DARS signal.

More complex to explain are the changes observed in the Raman spectra around 7 GPa (**Figures 8 and S8**). In α-GaGeTe, the Fermi resonance disappears above 7 GPa and several new modes appear close to those of α-GaGeTe (see pink symbols in **Figure 9a**) and with a similar pressure coefficient to those of α-GaGeTe. This result is a rather striking feature, as if a discontinuity in the pressure dependence of the modes' wavenumbers would occur around this pressure. This feature is also present for β-GaGeTe (see pink symbols in **Figure 9b**). The discontinuity is evident in the entire wavenumber range, but especially in the medium-wavenumber range (100-250 cm$^{-1}$). As a consequence, the $E_g^2$ and the $A_{1g}^2$ modes in α-GaGeTe get closer, so that eventually a mode crossing occurs around 13 GPa. Consequently, we attribute this discontinuity to the first phase transition observed in α-GaGeTe at approximately 7 GPa in the HP-XRD experiments, and we conclude that a similar phase transition occurs in β-GaGeTe.

With regard to the nature of the HP phase above 7 GPa in both polytypes, we have to consider that, despite the mentioned discontinuity, the number of experimentally detected modes does not change above the transition pressure. Additionally, for both polytypes it is evident that modes attributed to the low- and HP-phases coexist in a given pressure range, which is symptomatic of a first-order transition. The coexistence range is more extended for β-GaGeTe, suggesting a slower completion of the transition. It must be noted that in our theoretical calculations (**Figure 9a and 9b**) these discontinuities are not predicted, which can be considered a confirmation of the instability of the original phase above 7 GPa. Consequently, these results suggest that, for both polytypes, the observed Raman modes above 7 GPa can be explained by two hypotheses. On the one hand, these Raman modes may be attributable to the first HP phase, with a structure similar to that of the original phase, likely obtained by minor changes in the atomic positions or angles and/or to a very slight monoclinic distortion of the unit cell, undetectable in our XRD experiments, due to peak broadening and polytype coexistence. Alternatively, these modes may be attributable to the original low-pressure phase but distorted along the *c* axis, probably due to a collapse of the latter at the phase transition. In this way, the original and HP distorted phases would coexist in a certain pressure range near the onset of the transition and the Raman modes of the HP phase would not be detected because either the HP phase is metallic, with a negligible Raman scattering cross section (as it would occur for the proposed HP α'-GaGeTe structure, which has more Raman-active modes than α-GaGeTe (see **Table S8** in ESI)) or because the first HP phase actually has no Raman activity, such as the rocksalt-like phase related to the second HP phase. The same explanation given above for α-GaGeTe and its first HP phase can be applied to β-GaGeTe and its HP phase since the proposed HP phase β'-GaGeTe belongs to the same space group as β-GaGeTe, but with more Raman-active modes (see **Table S9** in ESI).

In summary, HP-RS measurements have allowed us: i) to assign most of the Raman-active modes of both polytypes of GaGeTe; ii) to calculate and discuss the pressure coefficients of the different modes in relation to the bonds present in the crystalline structure; and iii) to confirm the two phase transitions for both polytypes at pressures close to those determined by XRD experiments (near 7 and 15 GPa). The confirmation of any of the hypotheses given for the nature of the HP phases needs further experimental work, aimed at the determination of the two HP phases in both polytypes separately.

To conclude this section, we want to comment that we have performed complex calculations of the lattice thermal conductivity of bulk α-GaGeTe at several pressures, which are based on our lattice-dynamical calculations. Our calculations show that this polytype has much larger lattice thermal conductivities at HP (see **Figure S11** in ESI) than at RP (see Fig. 6 in Ref. [19]). The increase is larger in the direction along the *c* axis than along the *a* and *b* axes of the hexagonal unit cell. This can be understood by the much larger increase in the interlayer than in the intralayer interactions at low pressures in these layered compounds at HP as a consequence of the strong reduction of the interlayer space, so similar results (not shown) are reasonably expected for β-GaGeTe.

### 3.  Electronic properties

In the following, we explore in detail from the theoretical point of view the electronic band structure of both polytypes of GaGeTe and their behavior under compression. The value of the bandgap at RC has raised considerable controversies in the last years [7,9,12,18,36,72] and a recent work on α-GaGeTe showed that experimental measurements agree more with GGA-PBE calculations than with hybrid HSE06 calculations, all of them including spin-orbit coupling (SOC), since the latter show a large overestimation of the band gap [7]. In fact, our band structure calculations performed using a potential-only meta-GGA (MBJ) functional (see **Figure S12** in ESI) yield similar results to those of HSE06 functional, with a vastly overestimated direct bandgap at room pressure of 0.78 eV (0.59 eV), instead of an indirect bandgap of 8 meV (33 meV) for *α*-GaGeTe (*β*-GaGeTe) according to GGA-PBE and GGA-PBEsol calculations [7,19]. The change in bandgap character in HSE06 and MBJ calculations is due to the loss of valence and conduction band crossing as a consequence of the opening of the bandgap. For these reasons, we have calculated in this work the electronic band structure of α- and β-GaGeTe at different pressures with the GGA-PBEsol functional, including the SOC interaction, as we already did for the electronic band structure of both polytypes at room conditions in Ref. [19].

Regarding α-GaGeTe, the valence band (VB) shows a maximum (VBM) at RP along the Z-F direction, close to the Z point (hereafter Z' point), while the conduction band (CB) shows a minimum (CBM) along the U-Γ direction, close to the U point (hereafter U' point) (see **Figure 10a**). In addition, there are also two close CBMs at RP along the F-Γ direction close to the F point (hereafter F' point) and also at the Z-F direction close to the Z point (Z''). All in all, according to our calculations, bulk α-GaGeTe at RP is an indirect Z'-U' semiconductor (almost a semimetal) with a bandgap of 8 meV, another indirect Z'-Z bandgap of 54 meV, and with the direct bandgap (120 meV) located at the Z point. Regarding β-GaGeTe, the VBM at RP is along the Γ-K direction close to the Γ point (hereafter Γ' point), with another close VBM along the Γ-M direction, while the CBM at RP is at the Γ point, with another close CBM near the M point (see **Figure 11a**). Therefore, our calculations show that also *β*-GaGeTe at RP is an indirect Γ'-Γ semiconductor (almost a semimetal) with a bandgap of 33 meV and with the direct bandgap (61 meV) located at the Γ point.

With regard to the pressure dependence of the bandgaps of α-GaGeTe (**Figure 12a**), the VB dispersion along the Z-F direction and the CB along the U-Γ direction flatten as pressure increases, so the VBM moves to the Z point and the CBM moves slightly towards the U point, although a second CBM is also present close to the F point (see **Figure 10b,** corresponding to 2.3 GPa). In this way, the indirect Z'-Z indirect bandgap at RP disappears and the indirect Z'-U' bandgap transforms into the indirect Z-U' bandgap. On the other hand, the CBM approaches the VBM at HP so both indirect and direct bandgaps close as pressure increases. In fact, the indirect Z-U' bandgap closes at very small pressure (near RP) and the direct bandgap at Z closes at around 2.8 GPa. Above this pressure, the direct bandgap opens up linearly with increasing pressure, while the indirect bandgap continues closing (see **Figure 10c-d** corresponding to 5.4 and 7.5 GPa and **Figure 12a**), thus reinforcing the semimetallic character of α-GaGeTe at HP.

Regarding the pressure dependence of the bandgaps of *α*-GaGeTe (**Figure 12b**), the VB dispersion along the Γ-K direction and the CB along the Γ-M direction flatten as pressure increases, so the VBM moves to the Γ point and the CBM moves slightly towards the M point (see **Figure 11b,** corresponding to 2.8 GPa). This means that all indirect bandgaps become close to each other and end up in the Γ-M' bandgap at HP. On the other hand, the CBM approaches the VBM at HP so both indirect and direct bandgaps close with increasing pressure. In fact, the indirect Γ -M' bandgap closes around 1 GPa and the direct bandgap at Γ closes at around 2.8 GPa. Above this pressure, the direct bandgap opens up almost linearly with increasing pressure, while the indirect Γ-M' bandgap closes also linearly (see **Figures 11c-d** corresponding to 4.0 and 7.0 GPa and **Figure 12b**), thus reinforcing the semimetallic character of β-GaGeTe at HP. In summary, our calculations show that the direct and indirect bandgaps in β-GaGeTe evolve in a similar fashion as in α-GaGeTe, as can be seen by comparing **Figures 12a-b.**

Once the pressure dependence of the bandgaps of both polytypes has been analyzed, it is appropriate to compare the behavior of their bandgaps with those of related materials, such as III-VI semiconductors. The non-linear pressure dependence of the direct and indirect bandgaps near the Z and Γ points in α- and β-GaGeTe, respectively, at low pressures (below 3 GPa), as well as the linear pressure dependence of these bandgaps at high pressures (above 3.5 GPa) are similar to those observed for the direct and indirect bandgaps near the Z point in layered γ-InSe [73,74] and near the Γ point in layered ε-GaSe [74,75], respectively. In particular, the direct bandgap in γ-InSe (ε-GaSe) and in α-GaGeTe (β-GaGeTe) shows a slight decrease with increasing pressure up to 1.0 GPa (1.3 GPa) and 2.8 GPa (2.8 GPa), respectively, followed by a strong linear increase with increasing pressure up to the phase transition. On the other hand, the indirect bandgaps Z-B (Γ-M) and Z-U' (Γ-M') of γ-InSe (ε-GaSe) and α-GaGeTe (β-GaGeTe), respectively, show a strong decrease up to 1.0, 1.3, and 2.8 GPa (2.8 GPa), respectively, followed by a moderately steep linear decrease with increasing pressure, up to the phase transition. Moreover, the values of the direct and indirect bandgaps in both polytypes of GaGeTe are very similar at RP, as it is also the case for ε-GaSe. In fact, this semiconductor and both polytypes of GaGeTe exhibit an indirect bandgap at RP. The main difference is that the bandgaps in α- and β-GaGeTe are close to 0 eV, while those of ε-GaSe are close to 2 eV [75].

Despite the similarity of the direct and indirect bandgaps in ε-GaSe and in α- and β-GaGeTe, the change with pressure of the electronic band dispersion near the Z (Γ) point in α-GaGeTe (β-GaGeTe) is more similar to that of γ-InSe than to that of ε-GaSe. In particular, the evolution of the electronic band dispersion near the Z point in α-GaGeTe is just the opposite to that in γ-InSe. The latter semiconductor has the VBM at the Z point and a flat band dispersion along the Z-L direction at RP. As pressure increases, it develops a ring-shaped (or toroidal) VBM along the Z-L direction i.e., in the $k_z=0$ plane, with cylindrical symmetry with respect to the $k_z$ axis. Notably, this leads to a large continuous increase of the hole mobility along the Z-L direction in γ-InSe that doubles its value at RP near at 3.2 GPa [74]. Contrarily, a toroidal VBM along the Z-F direction (also close to the $k_z=0$ plane) occurs in α-GaGeTe at RP and it disappears at HP. This view leads us to consider that the opposite behavior is expected to occur in α-GaGeTe than in γ-InSe at HP i.e., the hole mobility will decrease severely with increasing pressure along the Z-F direction in α-GaGeTe, due to the severe increase in the hole effective mass in the layer plane as the toroidal VBM disappears.

For a better understanding of its features, the electronic band structure of bulk α-GaGeTe, can be compared more thoroughly with those of III-VI semiconductors. For that purpose, we have to consider that the electronic band structure of α-GaGeTe has an odd number of valence bands (13), similar to γ-InSe, and consequently has the direct bandgap around the Z point. In fact, the band structure of a monolayer of α-GaGeTe can be formed by the band structure of a monolayer of a III-VI semiconductor, like GaSe, if one breaks the Ga-Ga bonds inside layers and intercalates a germanene-like layer. Unlike in III-VI semiconductors, the presence of a germanene-like layer (with Ge-Ge bonds) leads to the appearance of two deep bands in the VB (between -7 and -12 eV from the VBM) that show very small dispersion along the Γ-Z direction. Additionally, the germanene-like layer does not contribute to the topmost VB; therefore, it does not affect the dispersion of the VB near the Z point. However, the germanene-like layer contributes significantly to the lowermost CB and seriously affects the dispersion of the CBM near the Z point since the CBM is mainly of Ge s character. This leads to a very flat dispersion along the Γ-Z direction and a strong dispersion along the Z-F direction (see **Figure 10a**). Both features indicate that the germanene-like layer acts as a 2D electronic band. The flat band dispersion in the Z-Γ direction involves a high density of states in the CBM, while the large band dispersion in the Z-F direction leads to a low electron effective mass and, most likely, to a high electron mobility in the layer plane.

An important difference of α-GaGeTe with respect to γ-InSe is the origin of the toroidal VBM in both compounds. The toroidal VBM occurs in γ-InSe due to the increase in the mixture of $p_z$ and p$_x$-p$_y$ orbitals with increasing pressure. Such a mixture is allowed by symmetry in γ-InSe (*R3m*, No. 160) because of the absence of a mirror plane perpendicular to the *c*-axis that is present in ε-GaSe (*P-6m2*, No. 187), where such a mixture is forbidden [74]. In α-GaGeTe, the centrosymmetric structure does not allow the mixture of the p$_z$ and p$_x$-p$_y$ orbitals, as in ε-GaSe. Consequently, the origin of the toroidal VBM in α-GaGeTe must be related to the band inversion, as in many topological insulators where the equal symmetry of the VBM and CBM along the Z-F direction produces anticrossings of both conduction and valence bands. Note that if both the valence and conduction bands have the same dispersion and they overlap, both bands should develop a dispersion with ring or M-shape. However, if there is an overlap of the topmost VBs and lowermost CBs and one band shows a much smaller dispersion than the other, as for the topmost VB of α-GaGeTe along the Z-F direction, the ring or M-shape is observed only in the band with the smaller dispersion. We claim that this is likely the origin of the ring-shape VBM in α-GaGeTe near the Z point. We want to stress that the same reasoning is valid for β-GaGeTe, where a ring or M-shape is observed in the VBM near Γ along the Γ-K direction at RP (**Figure 11a**). The ring shape is observed in β-GaGeTe due to the overlap of valence and conduction bands, while such a ring shape is not observed in ε-GaSe because there is no mixture of p$_z$ and p$_x$-p$_y$ orbitals and no overlap between valence and conduction bands.

On the other hand, the behavior of the electronic band structure of α-GaGeTe at HP can be understood as follows. Since the VBM around Z is composed of Te-p$_z$ orbitals and the CBM at Z is composed of Ge-s orbitals, with increasing pressure, in general, the CBM at Z tends to move to higher energies faster than the VBM near Z. Consequently, there is a net increase of the direct bandgap at Z under compression. In this context, the initial decrease in the direct bandgap at low pressures (**Figure 12a**) is due to the presence of interlayer forces related to the closing of the vdW gap, as it also happens in III-VI semiconductors (see discussion in Ref. [73]) On the other hand, the CBMs at the U' and F' points are mainly formed by Ga-*p* orbitals that blueshift at HP at smaller rates than Te-*p$_z$* orbitals in the VBM. The different behavior is due to the much smaller effect of pressure on Ga-Te and Ga-Ge bond distances than in Te-Te interlayer distances. Consequently, there is a net decrease of the indirect bandgap under compression that leads to an increasing semimetallic character in α-GaGeTe at HP. Again, the same reasoning given in this paragraph is valid to explain the behavior of the direct and indirect bandgaps (in this case related to the Γ and M' points) in β-GaGeTe at HP. In summary, the behavior of the direct and indirect bandgaps of α- and β-GaGeTe under compression are pretty similar to those found in layered γ-InSe and ε-GaSe, respectively [73,75], thus showing the similar nature of the electronic band structure in all these compounds. The main difference between both polytypes of GaGeTe and III-VI semiconductors is the presence of the internal germanene-like sublayer in GaGeTe polytypes, that contributes significantly to closing the bandgap in these materials. This makes them low-bandgap semiconductors, with semimetallic character and, consequentl,y topological properties at RP.

Finally, we want to briefly comment the effect of pressure on the topological properties of both polytypes of GaGeTe. In a previous work, a strong and weak topological character was confirmed for α-GaGeTe and β-GaGeTe, respectively, due to the presence of band inversion at the Z point (Γ point) of the BZ [19]. This band inversion causes the ring shape of the VBM in both polytypes, as already commented. Our calculations of topological invariants (see details in ref. [19]), show that both polytypes lose their topological properties as pressure increases. This result is consistent with the loss of the ring shape of the VBM already observed around 2.5 GPa in both polytypes (**Figures 10b and 11b**). Therefore, the loss of the topological properties in both polytypes at HP can be ascribed to the disappearance of the band inversion caused by the stronger energy increase of Ge-s states than of Te-$p_z$ states at HP, leading to the calculated net increase of the direct bandgap at HP.

## Conclusion

An exhaustive joint experimental and theoretical study of GaGeTe under compression has been performed. From the structural point of view, we have found that $\alpha$ and $\beta$ polytypes of GaGeTe have anisotropic compressibility and a low bulk modulus, around 40 GPa. Two phase transitions were found in the studied pressure range, from room pressure up to about 19 GPa. Structures with 6-fold coordination for Ga and Ge atoms and related to the tetradymite structure of $Bi_2Te_3$ have been proposed for the intermediate-pressure phase. On the other hand, a disordered and distorted rocksalt-like structure, also with 6-fold coordination for Ga and Ge atoms, is proposed for the highest-pressure phase, due to the simplification of the XRD patterns and the disappearance of the Raman activity above 15 GPa.

Raman spectroscopy under compression allowed the comparison of the experimental and theoretical phonon frequencies as a function of pressure and to discuss in detail the pressure coefficients and Grüneisen parameters of the Raman active modes of both polytypes with reference to their structural features. The softening of calculated low-frequency IR and silent modes for the $\alpha$ and $\beta$ polytype, respectively, supports the Ga-Ge bond-breaking mechanism that results in the proposed increase of the atomic coordination of Ga and Ge at the first phase transition.

Finally, the electronic band structure of both polytypes of GaGeTe and their pressure dependence has been discussed by theoretical calculations, which confirm that both polytypes are topological semimetals with very low band gaps at room pressure. Upon moderate compression the direct and indirect bandgaps decrease, but the direct bandgaps are found to open above a certain pressure. This behavior is similar to that observed in related layered compounds, such as $\gamma$-InSe and $\varepsilon$-GaSe. The increase of the direct bandgap at pressures above 2.5 and 3 GPa for α-GaGeTe and β-GaGeTe, respectively, causes the loss of the topological features. Above these pressures, thus, both polytypes become trivial insulators.

**Declaration of competing interest**

The authors declare that they have no known competing financial interests or personal relationships that could have appeared to influence the work reported in this paper.

**Data availability**

Data will be made available on request.

**Acknowledgements**

This publication is part of the project MALTA Consolider Team Network (RED2022-134388-T), financed by MINECO/AEI/10.13039/501100003329; by I+D+i projects PID2019-106383GB-41/42/43 and FIS2017-83295-P financed by MCIN/AEI/10.13039/501100011033; and by project PROMETEO CIPROM/2021/075 (GREENMAT), financed by Generalitat Valenciana. This study forms part of the Advanced Materials program and is supported by MCIN with funding from European Union NextGeneration EU (PRTR-C17.I1) and by Generalitat Valenciana through projects MFA/2022/007 and MFA/2022/025 (ARCANGEL). J.A.S. acknowledges the Ramon y Cajal fellowship (RYC-2015-17482) for financial support. E.B. acknowledges the Universitat Politècnica de València for his postdoctoral contract (program PAID-10-21). We thank V. Monteseguro (Universidad de Cantabria) and Dr. H. Osman (Universidad de Valencia) for the fruitful discussions on the subject. Authors thank the ALBA Synchrotron for the XRD experiments performed at the MSPD beamline with the collaboration of ALBA staff (proposals No. 2015021222 and 2020074395) as well as the Elettra Sincrotrone Trieste for XRD experiments performed at the XPRESS beamline (proposal 20215005, in-house experiment). Finally, the authors acknowledge the computing time provided by MALTA Cluster and Red Española de Supercomputación (RES). E.Ld.S. acknowledges the High Performance Computing Chair - a R&D infrastructure (based at the University of Évora; PI: M. Avillez), endorsed by Hewlett Packard Enterprise, and involving a consortium of higher education institutions, research centers, enterprises, and public/private organizations; and the Portuguese Foundation of Science and Technology for the financial support with the CEEC individual fellowship 5th edition, Reference 2022.00082.CEECIND.

# Tables

**Table 1:** Experimental and theoretical (in square brackets) axial compressibility for the α and β phases of GaGeTe under compression. Experimental values have been obtained by fitting the data of the three XRD experiments simultaneously.

| Phase | $\chi_a$ ($10^{-3}$ GPa$^{-1}$) | $\chi_b$ ($10^{-3}$ GPa$^{-1}$) |
|---|---|---|
| α | 6.09(5) [6.41] | 13.27(5) [12.37] |
| β | 5.61(5) [5.7] | 8.77(5) [13.86] |

**Table 2:** Experimental and theoretical (in parentheses) ambient pressure volume ($V_0$) and bulk modulus ($B_0$) for the α and β polytypes of GaGeTe under compression, as calculated from data in Figure 6. Similar pressure range for the experimental and theoretical values were used. Data from all experiments have been used in the calculations. To allow a meaningful comparison between different data sets, a 2$^{nd}$ order Birch-Murnaghan EOS has been used to fit all the data, thus the first derivative of the bulk modulus, $B_0$', has a fixed value of 4.

| Phase | α | | β | |
|---|---|---|---|---|
| | Theory | Experiment | Theory | Experiment |
| $V_0$ (Å$^3$) | 485(1.5) | 490.7(6) | 322.9(8) | 313.0(4) |
| $B_0$ (GPa) | 45(1.5) | 41.6(7) | 44(1.3) | 48.5(8) |

**Table 3.** Parameters of the second-order polynomial fit of the theoretical and experimental zero-pressure Raman frequencies, $\omega=\omega_0+aP+bP^2$, Grüneisen parameters, $\gamma$, and relative difference between experimental and theoretical frequencies ($R_\omega$) for α-GaGeTe. For the calculation of the theoretical and experimental Grüneisen parameters, bulk moduli of 41.6 and 45 GPa have been used, respectively (see main text).

| Mode | Theory | | | | Experiment | | | | |
|---|---|---|---|---|---|---|---|---|---|
| | $\omega_0$ (cm$^{-1}$) | $a$ (cm$^{-1}$/GPa) | $b$ (cm$^{-1}$/GPa$^2$) | $\gamma$ | $\omega_0$ (cm$^{-1}$) | $a$ (cm$^{-1}$/GPa) | $b$ (cm$^{-1}$/GPa) | $\gamma$ | $R_\omega$ (%) |
| $E_g^1$ | 40.2 | 1.44 | -0.035 | 1.61 | 41[a] | 1.4(4) | -0.02(4) | 1.42 | 1.66 |
| $A_{1g}^1$ | 77.3 | 4.86 | -0.10 | 2.83 | 79 | 5.2(4) | -0.21(5) | 2.78 | 1.75 |
| $E_g^2$ | 177.44 | 3.13 | -0.058 | 0.79 | 179 | 2.8(5) | 0.03(7) | 0.66 | 0.62 |
| $A_{1g}^2$ | 200.0 | 2.71 | -0.050 | 0.61 | 206 | 2.2(2) | -0.05(3) | 0.45 | 2.94 |
| $E_g^3$ | 276.0 | 5.18 | -0.07 | 0.80 | 280 | 6.0(4) | -0.20(5) | 0.89 | 1.30 |
| $A_{1g}^3$ | 285.5 | 5.62 | -0.10 | 0.89 | 295 | 4.8(3) | -0.09(4) | 0.68 | 3.39 |

[a] This mode is not visible at room conditions, due to the edge filter cut-off. As such, this value is obtained as the result of the quadratic fit of $\omega(P)$ for the experimental data.

**Table 4**. Parameters of the second-order polynomial fit of the theoretical and experimental zero-pressure Raman frequencies, $\omega=\omega_0+aP+bP^2$, Grüneisen parameters, $\gamma$, and relative difference between experimental and theoretical frequencies ($R_\omega$) for β-GaGeTe. For the calculation of the theoretical and experimental Grüneisen parameters, bulk moduli of 48.5 and 44 GPa have been used, respectively (see main text).

| Mode | Theory | | | | Experiment | | | | $R_\omega$ (%) |
|---|---|---|---|---|---|---|---|---|---|
| | $\omega_0$ (cm$^{-1}$) | $a$ (cm$^{-1}$/GPa) | $b$ (cm$^{-1}$/GPa$^2$) | $\gamma$ | $\omega_0$ (cm$^{-1}$) | $a$ (cm$^{-1}$/GPa) | $b$ (cm$^{-1}$/GPa) | $\gamma$ | |
| $E_2^1$ | 17.2 | 3.27 | -0.19 | 8.35 | - | - | - | - | - |
| $E_2^2$ | 35.6 | -0.36 | 0.02 | -0.45 | - | - | - | - | - |
| $E_1^1$ | 38.4 | 2.07 | -0.16 | 2.37 | 34 | 1.8(1) | -0.06(2) | 2.61 | 12.06 |
| $E_1^2$ | 53.7 | -1.25 | 0.10 | -1.02 | 59 | -0.70(15) | -0.02(2) | -0.57 | 9.74 |
| $E_2^3$ | 54.1 | -0.71 | 0.02 | -0.58 | - | - | - | - | - |
| $A_1^1$ | 76.7 | 4.47 | -0.10 | 2.56 | 65 | 4.93(1.4) | -0.17(2) | 3.67 | 16.31 |
| $A_1^2$ | 138.2 | 2.52 | -0.04 | 0.80 | 167 | 2.9(2) | -0.06(3) | 0.85 | 18.72 |
| $E_1^3$ | 173.0 | 3.41 | -0.08 | 0.87 | 178 | 3.1(3) | -0.05(4) | 0.83 | 3.09 |
| $E_2^4$ | 173.2 | 3.26 | -0.07 | 0.83 | - | - | - | - | - |
| $E_1^4$ | 174.9 | 4.45 | 0.00 | 1.12 | - | - | - | - | - |
| $E_2^5$ | 175.1 | 4.50 | -0.11 | 1.13 | - | - | - | - | - |
| $A_1^3$ | 198.8 | 2.87 | -0.07 | 0.63 | - | - | - | - | - |
| $A_1^4$ | 265.1 | 5.11 | -0.07 | 0.85 | 261 | 5.3(4) | -0.17(9) | 0.99 | 1.50 |
| $E_1^5$ | 268.4 | 5.60 | -0.10 | 0.92 | 285 | 5.4(2) | -0.13(3) | 0.92 | 5.89 |
| $E_2^6$ | 268.5 | 5.54 | -0.09 | 0.91 | - | - | - | - | - |
| $A_1^5$ | 283.1 | 5.56 | -0.11 | 0.86 | 294 | 5.0(2) | -0.11(3) | 0.82 | 3.91 |

# Figure Captions

**Figure 1**. General view of the crystal structure of a) α-GaGeTe and b) β-GaGeTe at RP. Each atom type is indicated, along with the coordination polyhedra: violet, germanium; green, gallium; yellow, tellurium. The S.G. and experimental (theoretical) cell parameters of α-GaGeTe are: S.G. $R\bar{3}m$ (No. 166, Z=6), $a$ = 4.048 Å (4.0495 Å), $c$ = 34.734 Å (34.4336 Å), $V$=492.91 Å$^3$ (489.01 Å$^3$) The S.G. and experimental (theoretical) cell parameters of β-GaGeTe are: S.G. $P6_3mc$ (No. 186, Z=6), $a$ = 4.0379 Å (4.04518 Å), $c$ = 22.1856 Å (23.00478 Å), $V$=313.27 Å$^3$ (326.00 Å$^3$). Atomic positions for both polytypes are assumed as in ref. [19].

**Figure 2**. XRD patterns of a pure α-GaGeTe sample as measured in the Alba-1 experiment (ALBA synchrotron, Spain, BL04-MSPD beamline, $\lambda$=0.4246 Å) at selected pressures, up to 13.6 GPa. Pressure, $P$ (in GPa), is indicated at the right of each pattern. Asterisks in the pattern at 5.2 GPa indicate features which change at 9.3 GPa as a symptom of a phase transition taking place above 5.2 GPa.

**Figure 3**. Experimental XRD patterns of mixed $α$-GaGeTe and $β$-GaGeTe samples as measured in the Alba-2 experiment at selected pressures (ALBA Synchrotron, Spain, BL04-MSPD beamline, $\lambda$=0.4642 Å), up to 19.6 GPa and on pressure release. Pressure, $P$ (in GPa), is indicated at the right of each pattern. Arrows indicate significant changes in the patterns, correlated with phase transitions. A red "r" above a pattern indicates data collected at pressure release.

**Figure 4**. Experimental XRD patterns of mixed $α$-GaGeTe and $β$-GaGeTe samples in the Elettra experiment at selected pressures (Elettra Sincrotrone Trieste synchrotron, Xpress beamline, $\lambda$=0.49585 Å), up to 19.6 GPa and on pressure release: (a) up to 11.9 GPa; (b) from 12.4 to 18.5 GPa, and (c) upon pressure release from 12.0 GPa to 10$^{-4}$ GPa (1 bar). Pressure, $P$ (in GPa), is indicated at the right of each pattern. Asterisks indicate significant changes in the patterns, correlated with phase transitions.

**Figure 5.** Experimental XRD pattern of GaGeTe at 15 GPa from the Elettra experiments (red circles).

**Figure 6.** Experimental and theoretical pressure dependence of structural parameters in α-GaGeTe (black) and β-GaGeTe (red): a) $a$ lattice parameter; b) $c$ lattice parameter; c), unit cell volume, $V$; and d) $c/a$ ratio. Symbols (lines) represent experimental (theoretical) data. The experimental data for each experiment are indicated with different symbols (see legend).

**Figure 7**. Theoretical pressure dependence of the thickness of the Ge-like sublayer for α- and β-GaGeTe (black circles and red squares, respectively). Red and black stripes for α- and β-GaGeTe respectively, are guides to the eye.

**Figure 8.** Raman spectra excited with the green laser ($\lambda$=532 nm) at selected pressures for: a)

α-GaGeTe and b) β-GaGeTe. Pressure, *P* (in GPa), is indicated besides of each spectrum. Asterisks above each spectrum indicate notable changes, as detailed in the main text. The lines show the evolution of the mode with the same color indicated besides to them.

**Figure 9.** Pressure dependence of the Raman-active frequencies of a) α-GaGeTe and b) β-GaGeTe. Open symbols indicate Raman-active modes detected experimentally with the red laser (λ=632.8 nm), closed symbols indicate modes detected experimentally with the green laser (λ=532 nm). The symmetry of each mode is indicated in color. Theoretically calculated modes are shown as lines with the same color code. In both plots, open and full violet squares represent the modes of the unknown HP phases as measured with the red and the green laser, respectively, after the first HP phase transition.

**Figure 10.** Theoretical (PBEsol+SOC) electronic band structure of α-GaGeTe along the U-Γ-Z-F-Γ-L directions at (a), 0 GPa; (b), 2.6 GPa; (c), 4.2 GPa and (d), 7.1 GPa.

**Figure 11.** Theoretical (PBEsol+SOC) electronic band structure of β-GaGeTe along the Γ-M-K-Γ-A-L-H-A-L-M-H-K directions at (a), 0 GPa; (b), 2.8 GPa; (c), 5.8 GPa and (d), 7.8 GPa. Given the short distance between the A, L and M, H points of the BZ of *β*-GaGeTe, these have been indicated as A|L (in violet) and M|H (in blue), for better readability. On the plots, the patterns corresponding to A|L and M|H are indicated in violet and in blue, respectively.

**Figure 12.** Pressure dependence of the theoretical direct (Z) and indirect (U-Z) bandgaps in α-GaGeTe (a) and of the direct and indirect bandgaps in β-GaGeTe (b).

# Figures

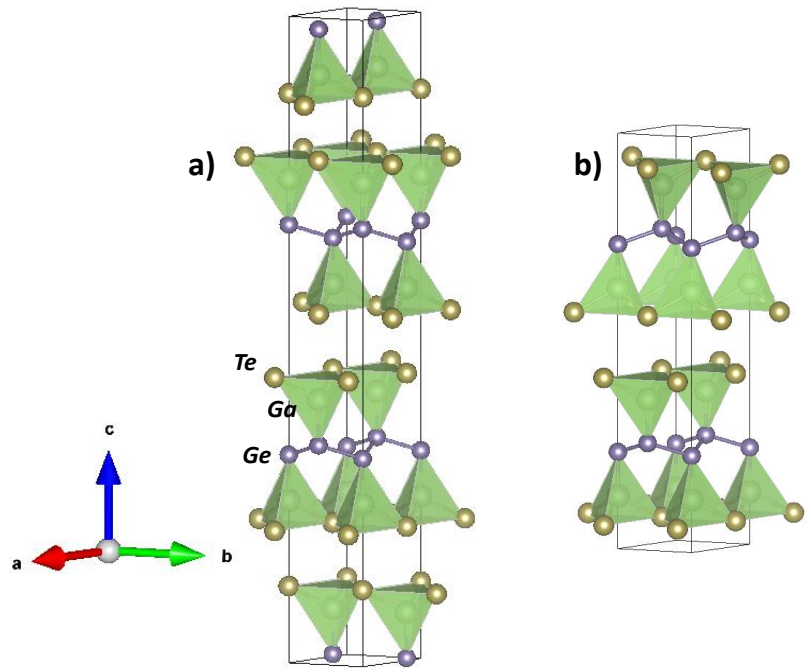

**Figure 1**

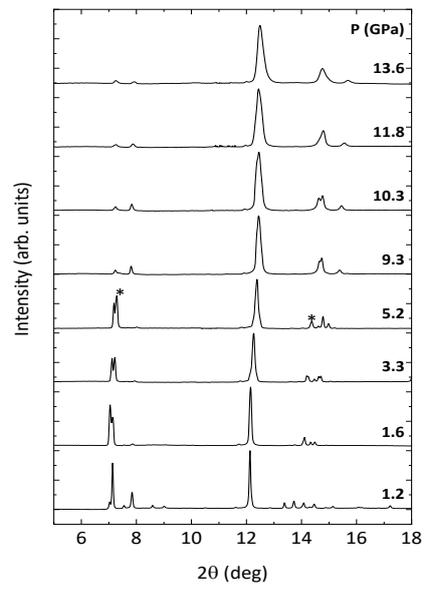

**Figure 2**

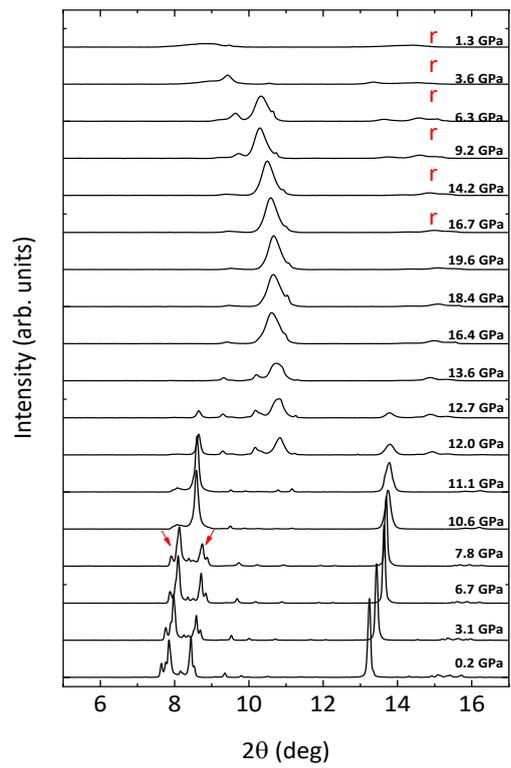

**Figure 3**

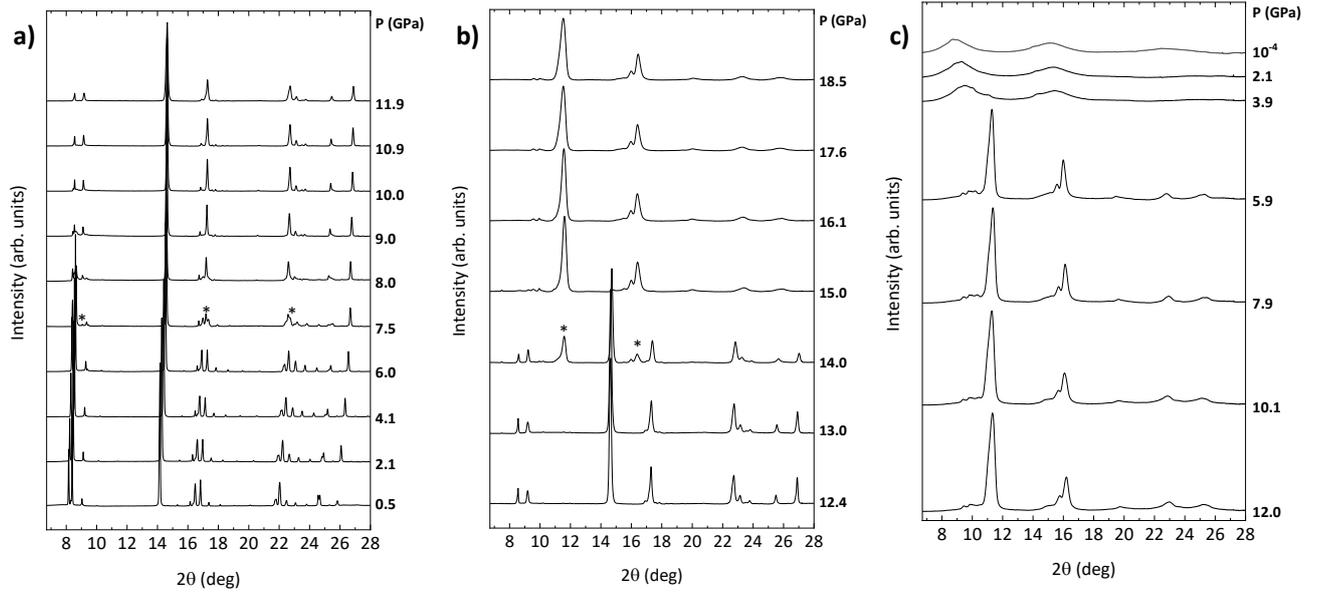

**Figure 4**

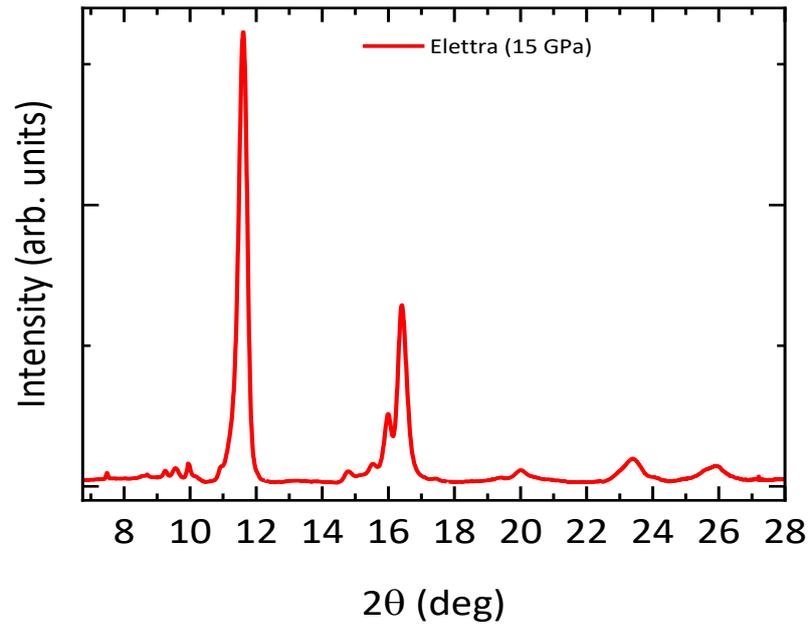

**Figure 5.**

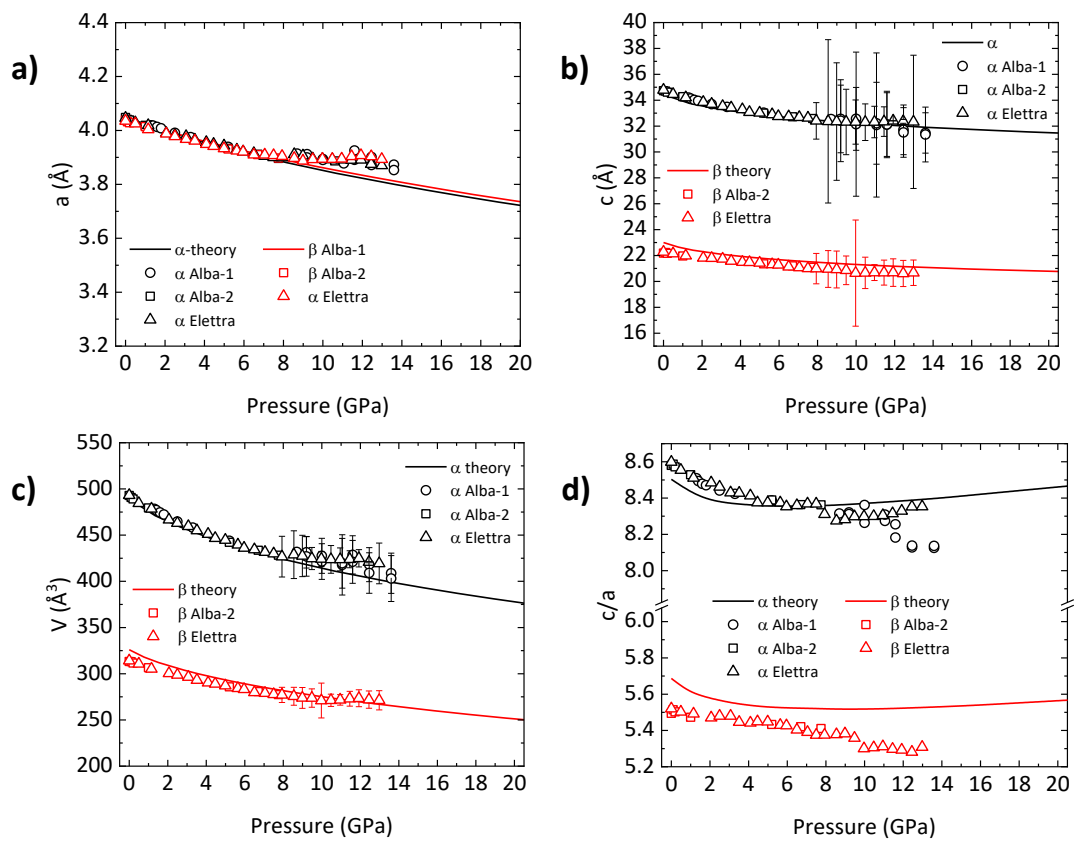

**Figure 6.**

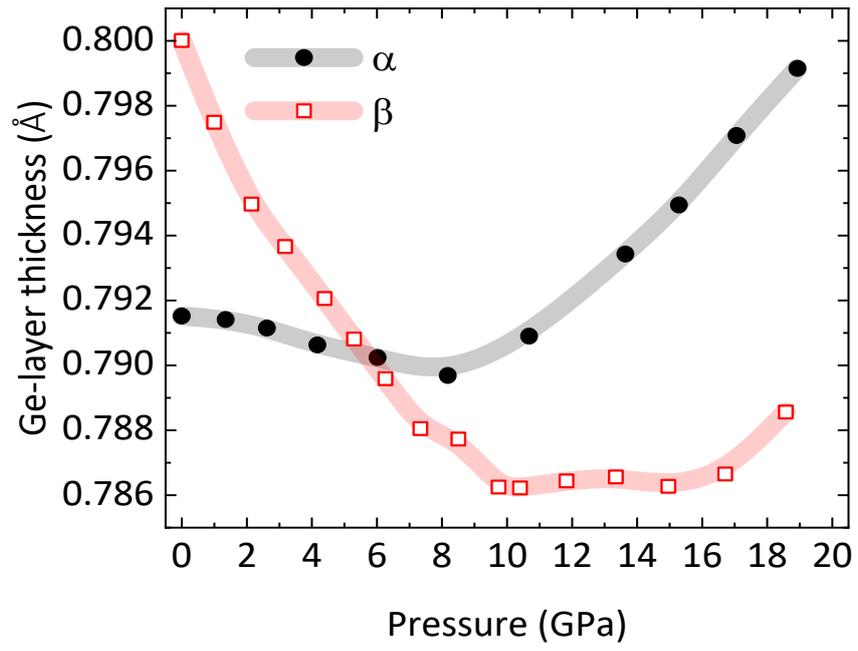

**Figure 7.**

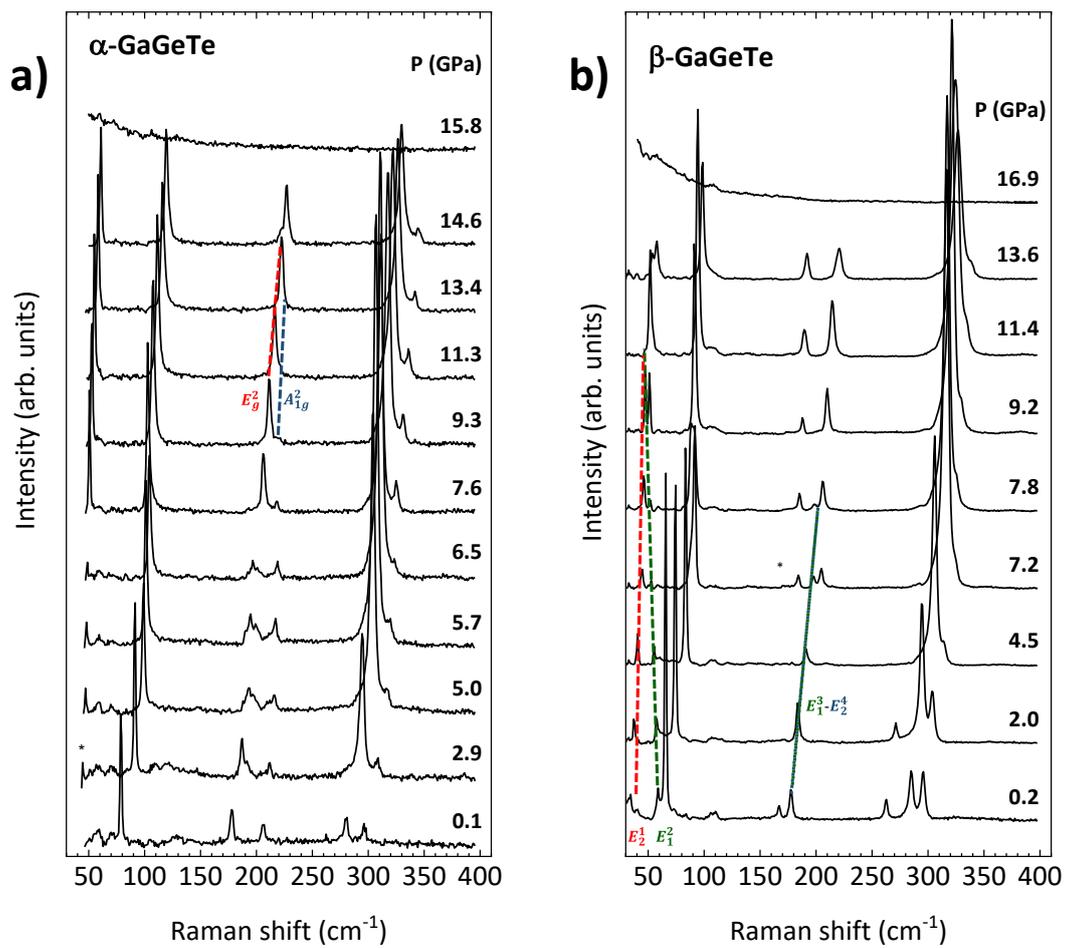

**Figure 8.**

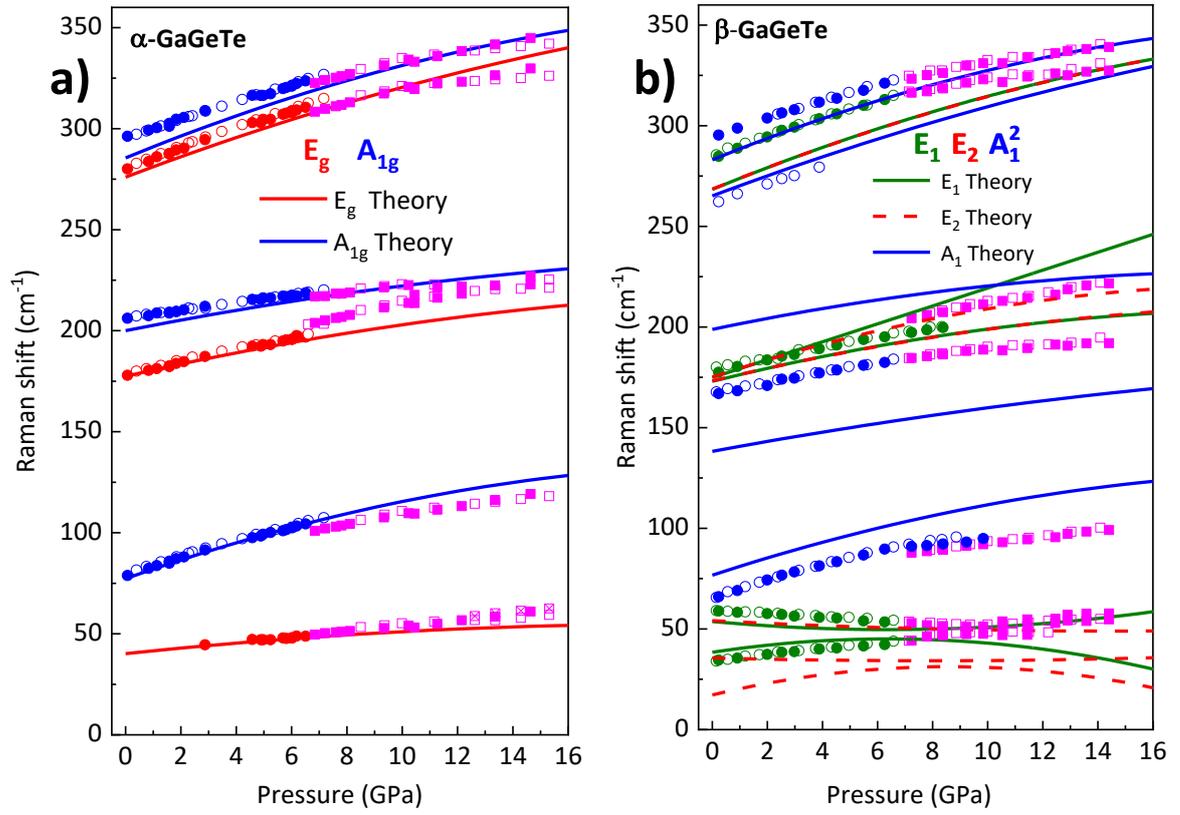

**Figure 9.**

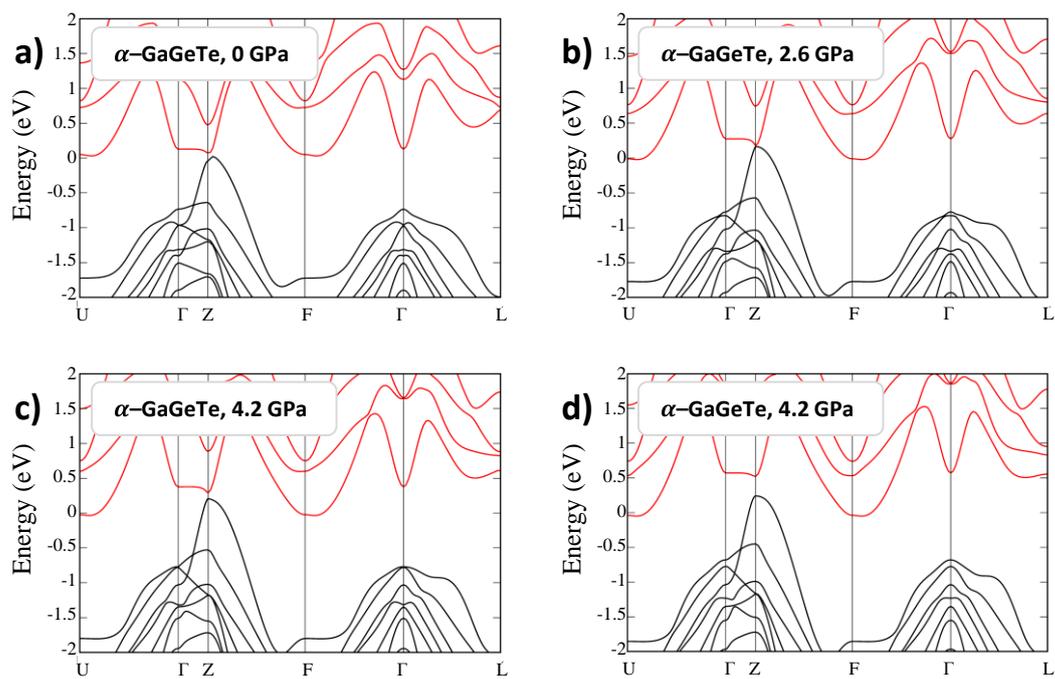

**Figure 10.**

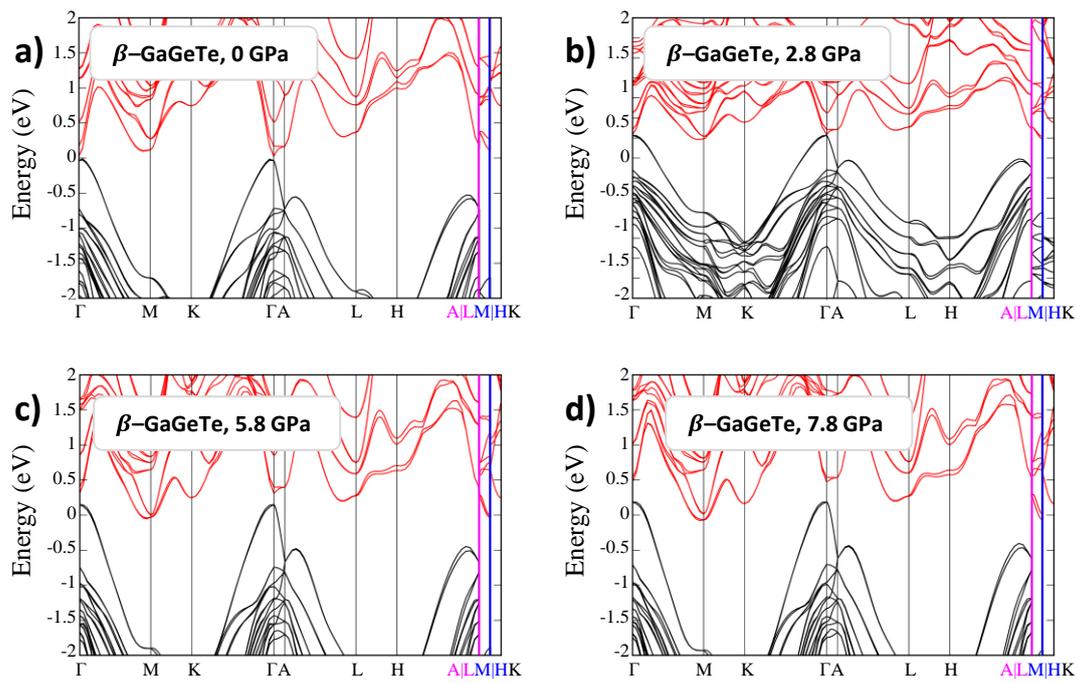

**Figure 11.**

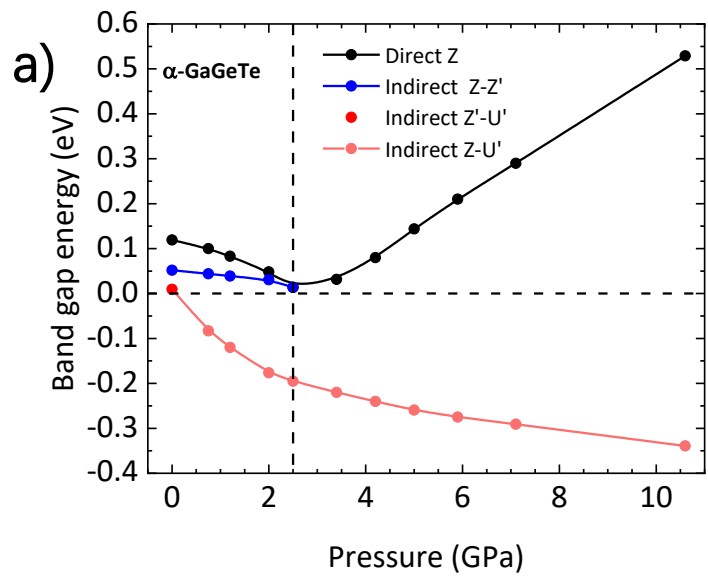

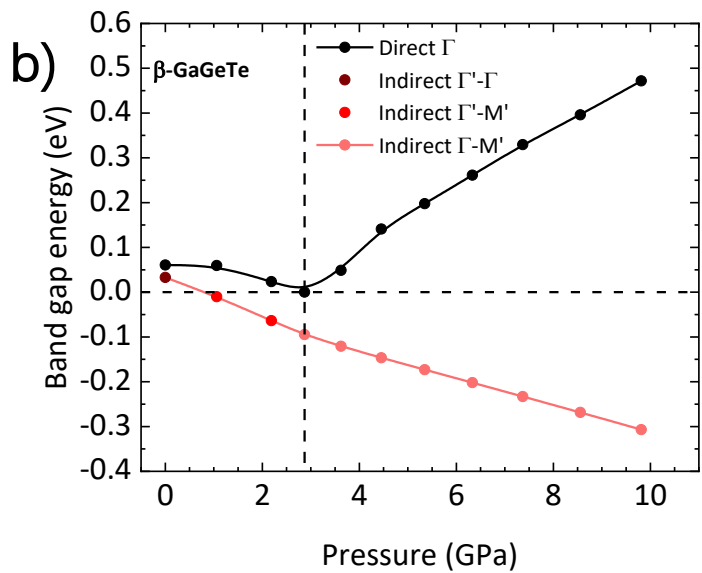

**Figure 12.**

# Supplementary Material of

# Structural, Vibrational, and Electronic Behavior of Two GaGeTe Polytypes under Compression.

E. Bandiello[1,*], S. Gallego-Parra[1,*], A. Liang,[2] J.A. Sans[1], V. Cuenca-Gotor[1], E. Lora da Silva[3], R. Vilaplana[4], P. Rodríguez-Hernández[5], A. Muñoz[5], D. Diaz-Anichtchenko[2], C. Popescu[6], F.G. Alabarse[7], C. Rudamas[8], C. Drasar[9], A. Segura[2], D. Errandonea[2], and F.J. Manjón[1]

[1] *Instituto de Diseño para la Fabricación y Producción Automatizada, MALTA Consolider Team, Universitat Politècnica de València, 46022 Valencia, Spain*

[2] *Departamento de Física Aplicada-ICMUV, MALTA Consolider Team, Universitat de València, 46100 Burjassot, Spain*

[3] *IFIMUP, Institute of Physics for Advanced Materials, Nanotechnology and Photonics, Department of Physics and Astronomy, Faculty of Sciences, University of Porto, Rua do Campo Alegre, 687, 4169-007 Porto, Portugal*

[4] *Centro de Tecnologías Físicas, MALTA Consolider Team, Universitat Politècnica de València, 46022 València, Spain*

[5] *Departamento de Física, Instituto de Materiales y Nanotecnología, MALTA Consolider Team, Universidad de La Laguna, La Laguna, 38205 Tenerife, Spain*

[6] *CELLS-ALBA Synchrotron Light Facility, MALTA Consolider Team, 08290 Cerdanyola del Vallès, Barcelona, Spain*

[7] *Elettra Sincrotrone Trieste, S.S. 14 - km 163,5 in AREA Science Park, 34149 Basovizza, Trieste, Italy*

[8] *Escuela de Física, Facultad de Ciencias Naturales y Matemática, Universidad de El Salvador, San Salvador, El Salvador*

[9] *Faculty of Chemical Technology, University of Pardubice, Pardubice 532 10, Czech Republic*

# 1.- Structural data

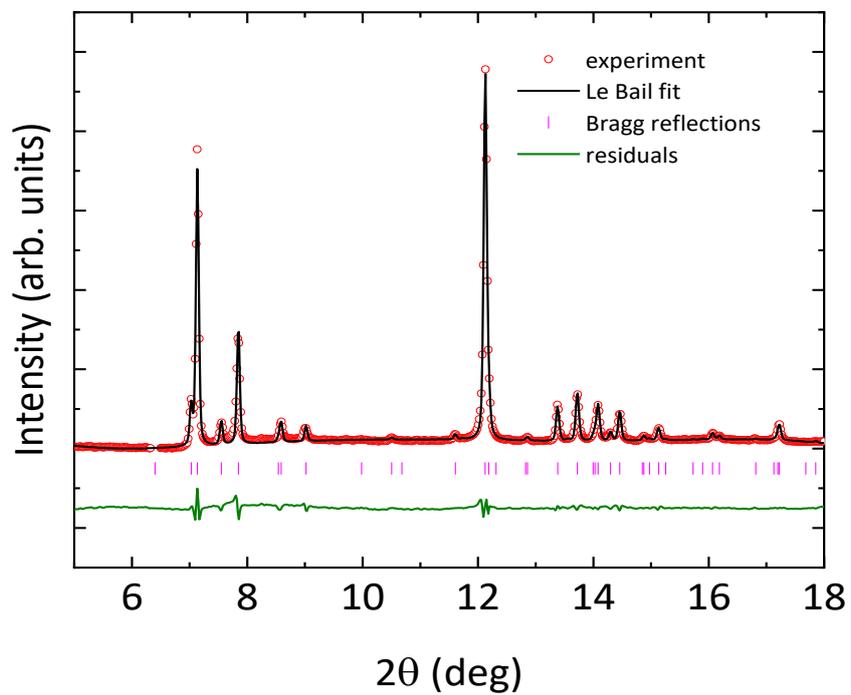

**Figure S1**. Experimental XRD pattern of α-GaGeTe at 1.2 GPa in the Alba-1 experiment (red circles, λ=0.4246 Å) and Le Bail refinement of the XRD pattern (black line). Magenta ticks show the position of Bragg reflections, while the green line represents the residuals of the Le Bail fit (difference between experimental and computer XRD patterns).

**Table S1.** Experimental and theoretical (PBEsol) unit-cell parameters for α- and β-GaGeTe at near RP pressure, except for the experimental cell parameters of the α polytype in the Alba-1 experiment that are shown at the minimum pressure of *P*=1.2 GPa.

| Polytype | | Experiment | | | | Theory |
|---|---|---|---|---|---|---|
| | | Alba-1 (1.2 GPa) | Alba-2 | Elettra | Ref [1] | (PBESol) |
| α $(R\bar{3}m)$ | *a* (Å) | 3.9924(1) | 4.0454(4) | 4.0459(1) | 4.0480 | 4.0495 |
| | *c* (Å) | 34.592(2) | 34.648(8) | 34.789(2) | 34.7340 | 34.4336 |
| | *V* (Å³) | 477.50(5) | 491.06(5) | 493.18(5) | 492.91 | 489.01 |
| β $(P6_3mc)$ | *a* (Å) | - | 4.0320(5) | 4.0350(1) | 4.0379 | 4.04518 |
| | *c* (Å) | - | 22.2334(5) | 22.274(5) | 22.1856 | 23.00478 |
| | *V* (Å³) | - | 313.024(5) | 314.08(5) | 313.27 | 326.00 |

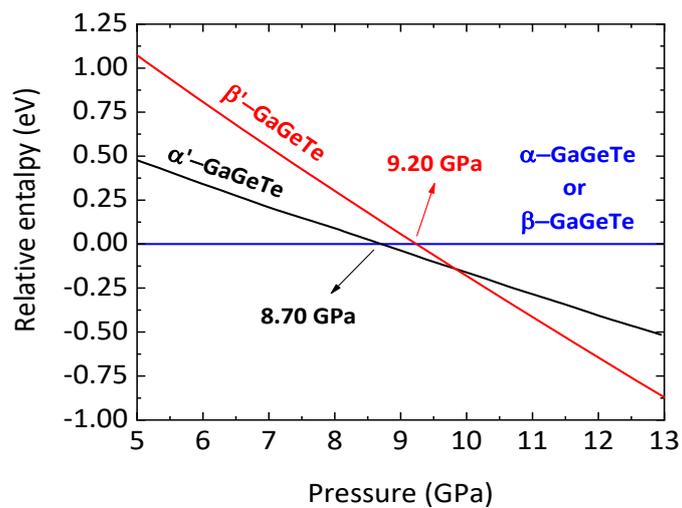

**Figure S2:** Theoretical relative enthalpy of the proposed high-pressure phases α'-GaGeTe (black line) and β'-GaGeTe (red line) with respect to those of the low-pressure phases α-GaGeTe and β-GaGeTe (blue line), respectively. The high-pressure phases become competitive above 8.70 GPa (α'-GaGeTe) and 9.20 GPa (β'-GaGeTe).

**Table S2.** Theoretical (PBEsol) atomic coordinates of the structure of α'-GaGeTe (s.g. $R\bar{3}m$, No. 166) at approximately 10 GPa. The theoretical PBEsol lattice parameters of the hexagonal unit cell are: $a$ = 3.6540 Å; $c$ = 30.6151 Å, and $V_0$ = 354.00 Å$^3$.

| Atom | Wyckoff site | x | y | z |
|---|---|---|---|---|
| Ga | 6c | 0 | 0 | -0.07287 |
| Ge | 6c | 0 | 0 | 0.64364 |
| Te | 6c | 0 | 0 | 0.20602 |

**Table S3.** Theoretical (PBEsol) atomic coordinates of the structure of β'-GaGeTe (s.g. $P6_3mc$, No. 186) at approximately 10 GPa. The theoretical lattice parameters of the hexagonal unit cell are $a = 3.6730$ Å, $c = 20.3705$ Å, and $V_0 = 238.00$ Å$^3$.

| Atom | Wyckoff site | x | y | z |
|---|---|---|---|---|
| Ga1 | 2b | 1/3 | 2/3 | 0.64062 |
| Ga2 | 2b | 1/3 | 2/3 | 0.3729 |
| Ge1 | 2a | 0 | 0 | 0.21627 |
| Ge2 | 2b | 1/3 | 2/3 | 0.28539 |
| Te1 | 2a | 0 | 0 | 0.44202 |
| Te2 | 2b | 1/3 | 2/3 | 0.06045 |

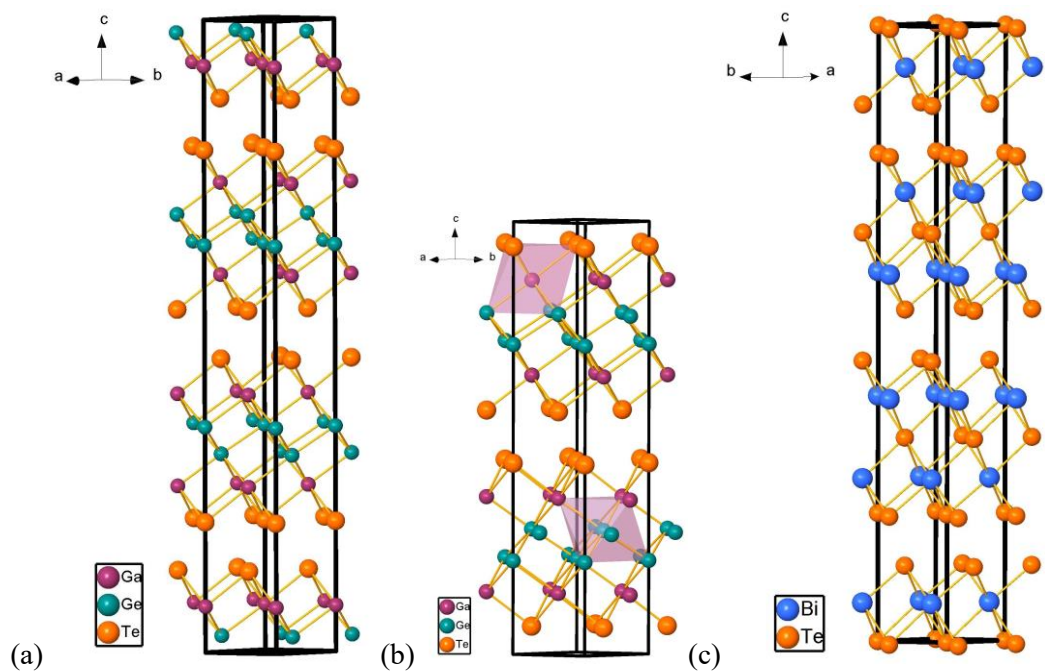

**Figure S3.** 3D view of the hexagonal unit cell of α'-GaGeTe (a) β'-GaGeTe (b), and Bi$_2$Te$_3$ (c). Ga, Ge, and Te atoms are represented in purple, green, and orange, respectively. The sixfold coordination of Ga and Ge atoms in both structures by the pink octahedra in the structure of β'-GaGeTe. Note that each monolayer of α'-GaGeTe and β'-GaGeTe is similar to that of the monolayer of Bi$_2$Te$_3$ where the central sublayer of Te is substituted by the germanene sublayer.

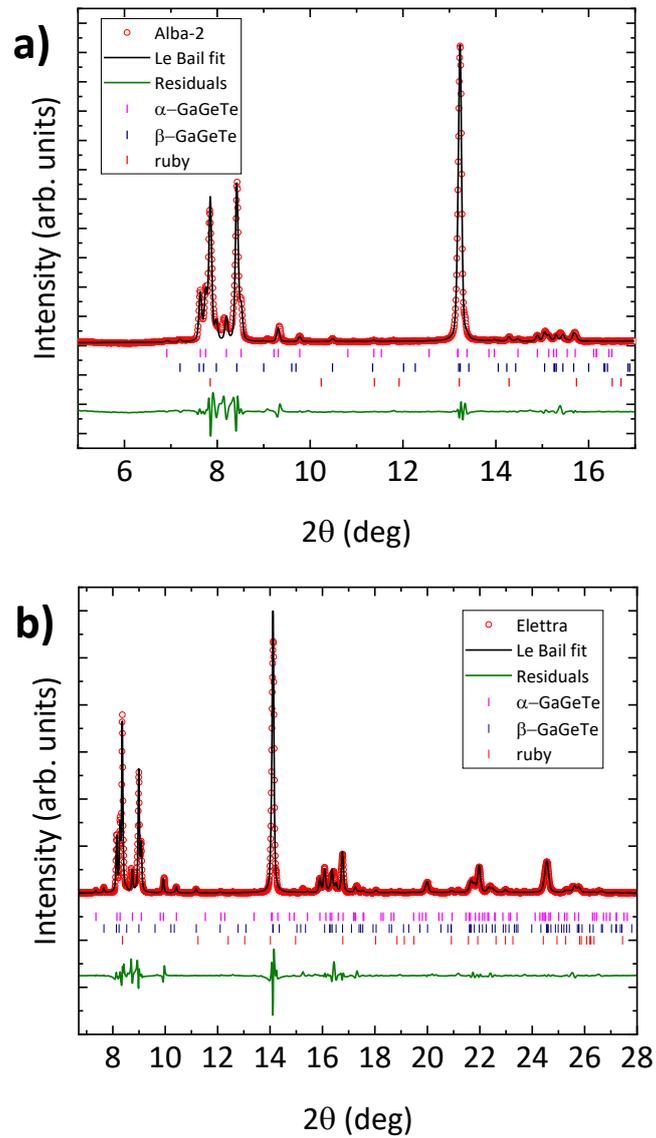

**Figure S4.** Experimental XRD data (red circles) and Le Bail fits (black line) for the (a) Alba-2, and (b) Elettra experiments. In both plots, residuals are represented by a green line. Bragg reflections for α-GaGeTe, β-GaGeTe, and ruby are shown (magenta, blue and red ticks, respectively).

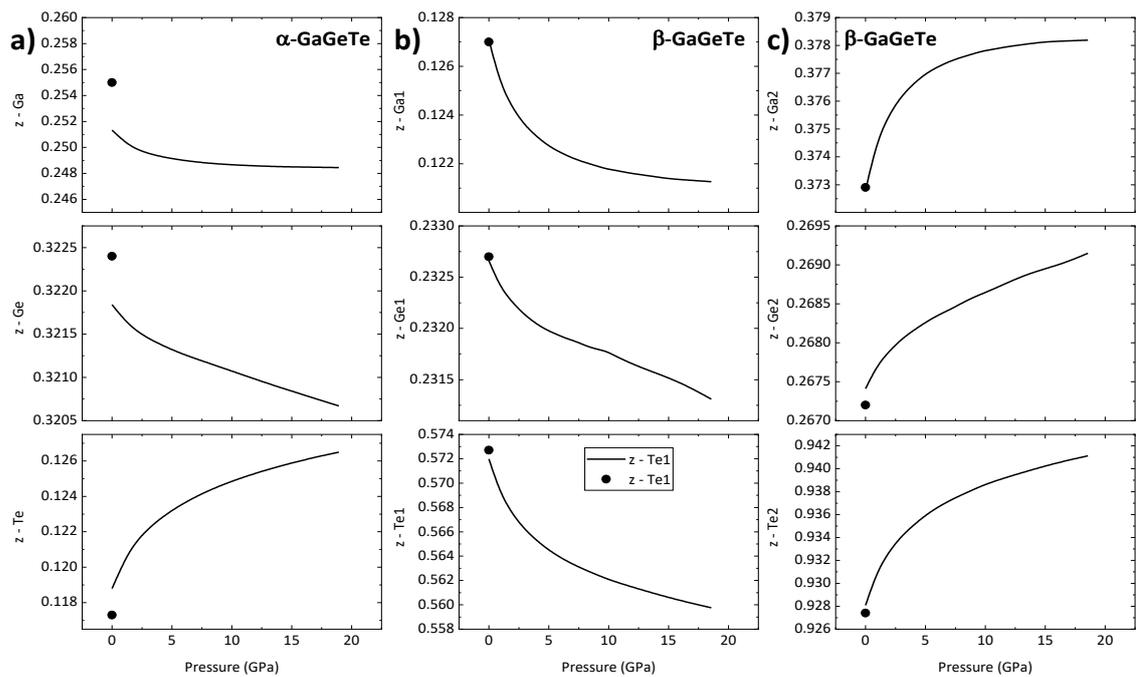

**Figure S5.** Theoretical pressure dependence of the *z* free atomic parameters of Ga, Ge and Te for a) α-GaGeTe and b) and c) β-GaGeTe. For comparison, experimental values at 0 GPa from Ref. [1] are plotted as symbols.

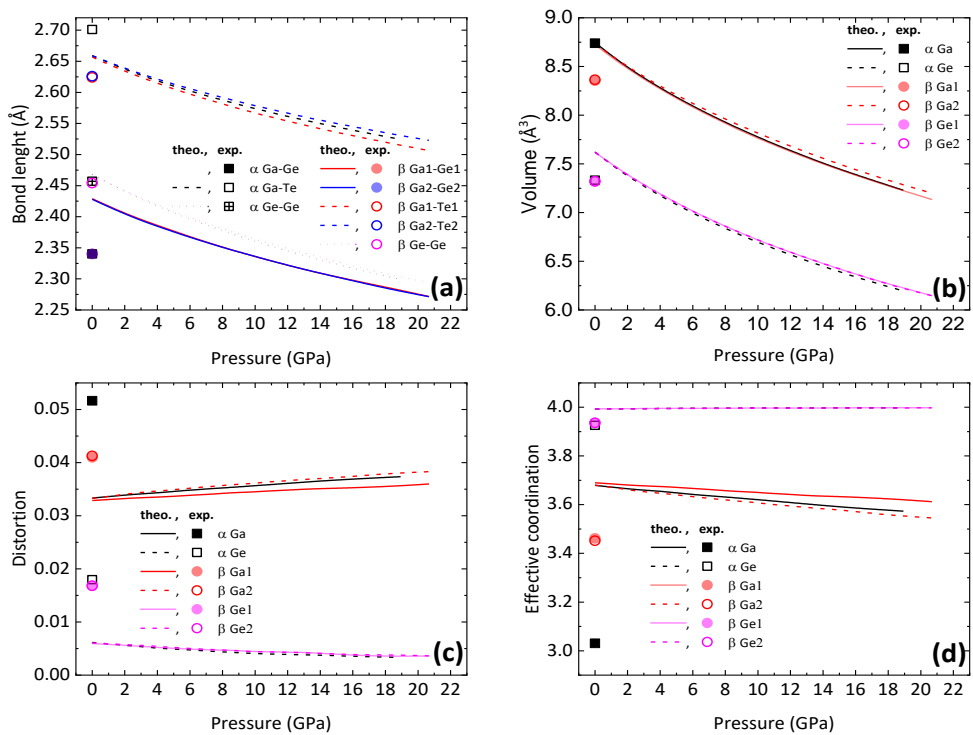

**Figure S6.** Theoretical (lines) pressure dependence of the bond lengths (a), polyhedral volumes (b), distortion index (c), and effective coordination (d) of Ga and Ge tetrahedra in α- (black lines and black symbols) and β-GaGeTe (red, blue and magenta symbols), respectively. Corresponding experimental values at room pressure, as measured from the data in Ref. [1], are shown as symbols. See legends for details. Ge-Ge distances are identical for Ge1 and Ge2 tetrahedra in β-GaGeTe. Additionally, due to their small experimental and theoretical differences, Ga-Ge distance in α-GaGeTe and the Ga1-Ge1 and Ga2-Ge2 distances in β-GaGeTe are graphically very close in panel (a), thus theoretical lines and experimental values at room pressure are superposed, respectively. The same happens for the volumes of Ge1 and Ge2 tetrahedra in β-GaGeTe in panel (b). For data with (partial) superposition, transparency was used (i.e., β Ga1-Ge1, β Ga2-Ge2 and α Ga-Ge in panel (a); β Ga1 (Ge1), and β Ga2 (Ge2) in panels (b), (c) and (d)). The low experimental value of the effective coordination for the Ga polyhedron in α-GaGeTe (panel (d)) is due to the fact that the calculated Ga-Te and Ga-Ge bond distances for this polyhedron are more similar to each other in theoretical calculations than in experimental data (see data in Table S5 for a numerical comparison).

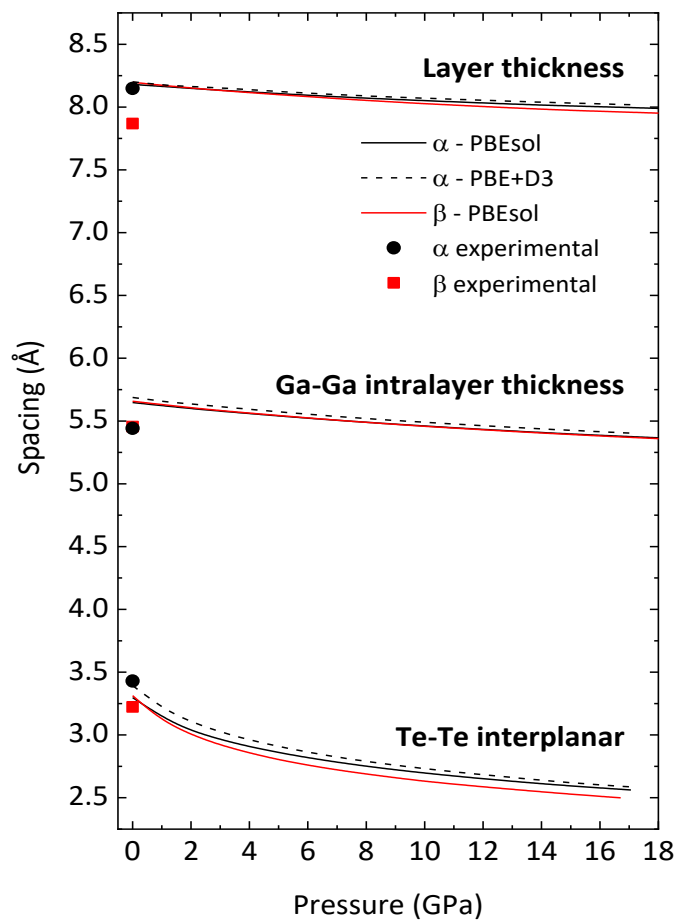

**Figure S7.** Theoretical (lines) pressure dependence of the thickness of a GaGeTe monolayer (layer thickness), interplanar distance between Ga planes inside a monolayer (Ga-Ga intralayer thickness) and interplanar distance (Te-Te interplanar) for α-GaGeTe and β-GaGeTe (in black and red, respectively). Experimental values of the corresponding magnitudes at room pressure, as reported in Ref. [1], are represented by black circles and red squares (α-GaGeTe and β-GaGeTe, respectively). Dashed black lines represent PBE+D3 calculations for α-GaGeTe and are shown to confirm the qualitative equivalence of PBEsol and PBE+D3 calculations, as mentioned in the main manuscript. Note that the calculated Ga-Ga intralayer thickness values for α and β-GaGeTe are almost coincident in all the pressure range, thus the corresponding lines are graphically superposed.

**Table S4.** Theoretical bond lengths at 0 GPa for the different Ga (upper part of the Table) and Ge (lower part of the Table) tetrahedra in α-GaGeTe and in β-GaGeTe. For each polytype, the three Ga-Te bonds of the Ga tetrahedra are identical, as are the Ge-Ge bonds of the Ge tetrahedra. For comparison, experimental values (in italics) are shown as well, below each theoretical value.

|  | **Ga1-Ge1** (Å) | (3×) **Ga1-Te1** (Å) | **Ga2-Ge2** (Å) | (3×) **Ga2-Te2** (Å) |
|---|---|---|---|---|
| α (Ga) | 2.42791 | 2.65906 | - | - |
|  | *2.34(3)* | *2.701(13)* | - | - |
| β (Ga1, Ga2) | 2.42902 | 2.65686 | 2.42810 | 2.65895 |
|  | *2.34(3)* | *2.624(13)* | *2.34(3)* | *2.626(14)* |
|  | **Ge1-Ga1** (Å) | (3×) **Ge1-Ge1** (Å) | (3×) **Ge2-Ge2** (Å) | **Ge2-Ga2** (Å) |
| α (Ge) | 2.42791 | 2.46831 | - | - |
|  | *2.34(3)* | *2.457(11)* | - | - |
| β (Ge1) | 2.42902 | 2.46787 | - | - |
|  | *2.34(3)* | *2.454(9)* | - | - |
| β (Ge2) | - | - | 2.46787 | 2.42810 |
|  | - | - | *2.454(9)* | *2.34(3)* |

**Table S5.** Theoretical bulk modulus $B_0$ of the different Ga and Ge tetrahedra in α-GaGeTe and β-GaGeTe, as calculated from data of Figure S6b using a 2$^{nd}$ order Birch-Murnaghan equation of state ($B_0$' fixed to 4).

|         | $V_0$ (Å$^3$) | $B_0$ (GPa) | $B_0$' |
|---------|---------------|-------------|--------|
| α (Ga)  | 8.728(4)      | 68.7(4)     | 4      |
| α (Ge)  | 7.606(5)      | 60.6(3)     | 4      |
| β (Ga1) | 8.712(4)      | 69.2(3)     | 4      |
| β (Ga2) | 8.723(2)      | 73.11(17)   | 4      |
| β (Ge1) | 7.611(2)      | 62.75(18)   | 4      |
| β (Ge2) | 7.609(2)      | 62.81(18)   | 4      |

## 2.- Vibrational data

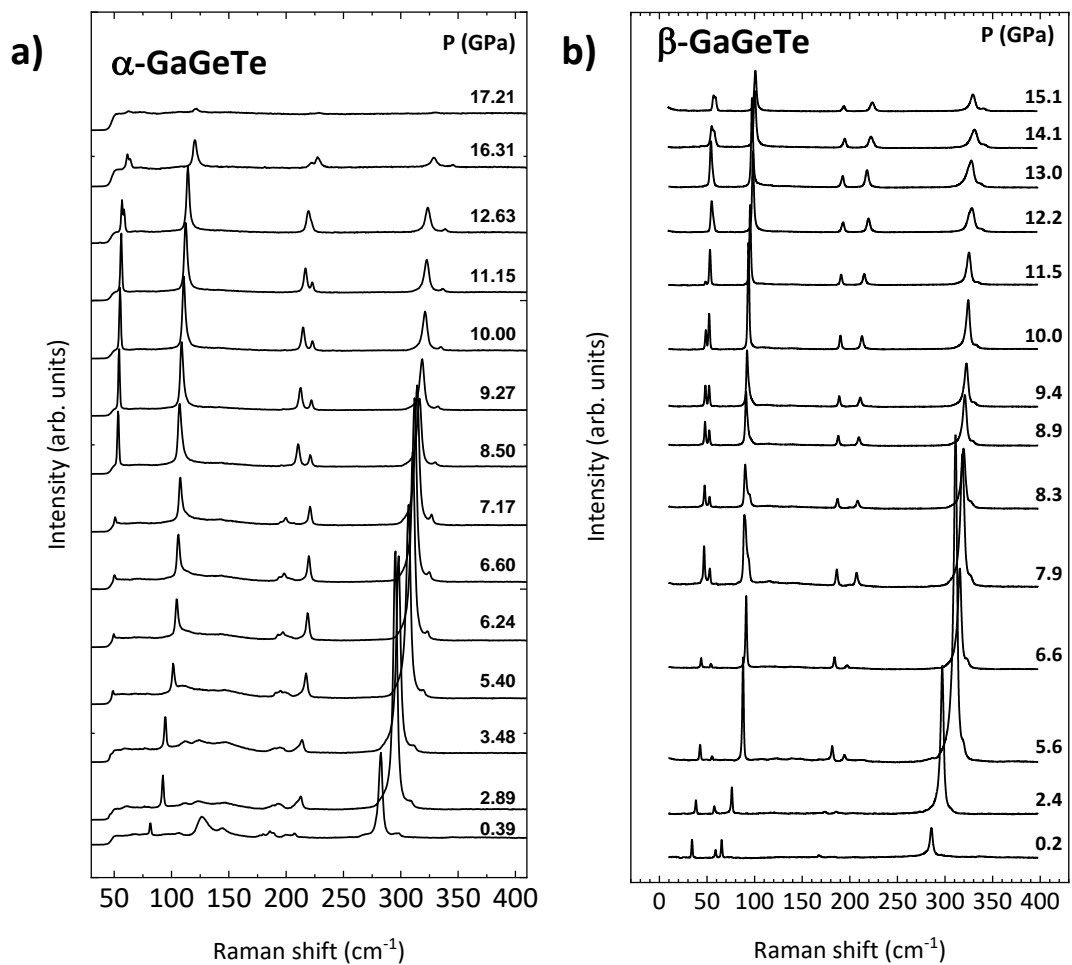

**Figure S8.** Raman spectra for a) α-GaGeTe and, b) β-GaGeTe at selected pressure (*P*), as obtained using a red laser (λ=632.8 nm). Pressure is indicated in GPa for each spectrum.

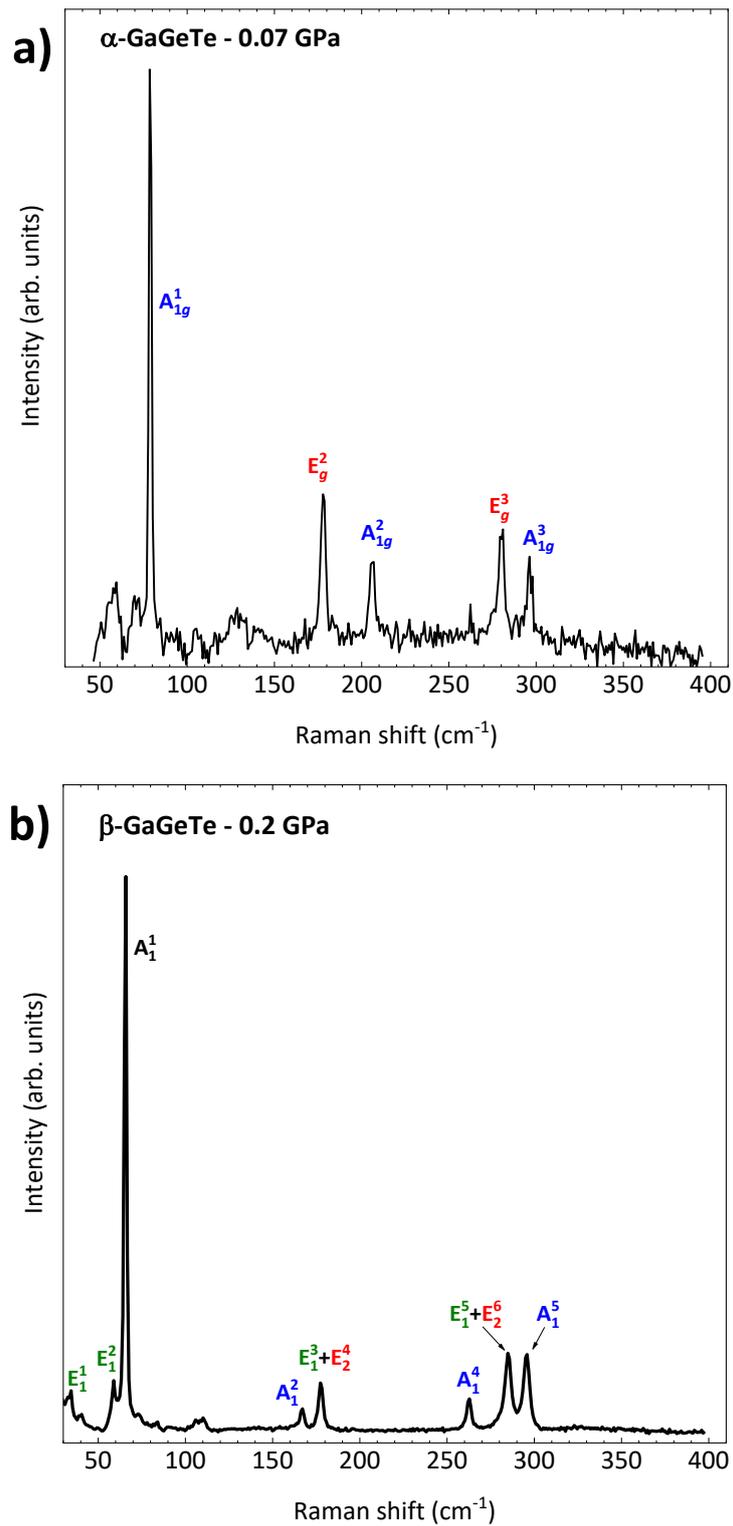

**Figure S9:** Raman spectra at nearly-room pressure for a) α-GaGeTe and, b) β-GaGeTe as measured with a green laser (λ=532 nm). The peaks corresponding to the Raman modes detected experimentally have been identified by their symmetry (according to Ref. [1]).

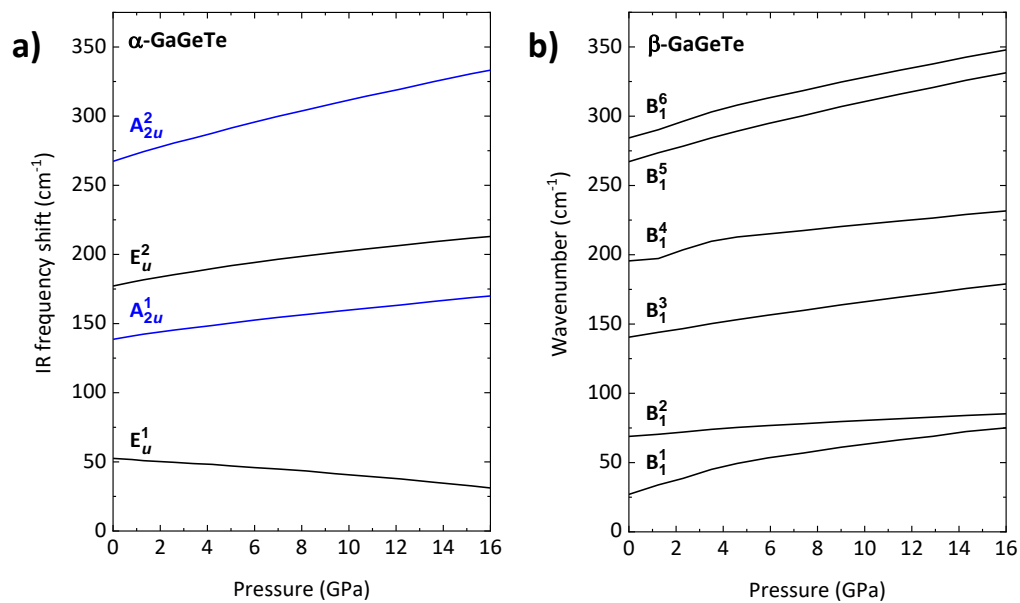

**Figure S10**. Theoretical (PBEsol) pressure dependence of the wavenumbers of IR-active modes in α-GaGeTe (a) and silent modes in β-GaGeTe (b). E and A modes in α-GaGeTe are plotted in black and blue respectively.

**Table S6.** Theoretical zero-pressure frequencies, pressure coefficients, and Grüneisen parameters, γ, obtained from a second-order polynomial fit, $\omega=\omega_0+aP+bP^2$, for the IR-active modes in α-GaGeTe. For the calculation of the theoretical Grüneisen parameters, a bulk modulus of 41.6 GPa has been used, respectively (see main manuscript).

| Mode | $\omega_0$ (cm$^{-1}$) | $a$ (cm$^{-1}$/GPa) | $b$ (cm$^{-1}$/GPa$^2$) | γ |
|---|---|---|---|---|
| $E_u^1$ | 52.3 | -0.9 | -0.03 | -0.72 |
| $A_{2u}^1$ | 139.1 | 2.4 | -0.03 | 0.72 |
| $E_u^2$ | 177.6 | 3.1 | -0.05 | 0.73 |
| $A_{2u}^2$ | 267.8 | 4.9 | -0.05 | 0.76 |

**Table S7**. Theoretical zero-pressure frequencies, pressure coefficients, and Grüneisen parameters, γ, obtained from a second-order polynomial fit, $\omega=\omega_0+aP+bP^2$, for the silent modes in β-GaGeTe. For the calculation of the theoretical Grüneisen parameters, a bulk modulus of 48.5 GPa has been used (see main text).

| Mode | $\omega_0$ (cm$^{-1}$) | $a$ (cm$^{-1}$/GPa) | $b$ (cm$^{-1}$/GPa$^2$) | $\Gamma$ |
|---|---|---|---|---|
| $B_1^1$ | 28.2 | 4.9 | -0.13 | 8.43 |
| $B_1^2$ | 68.9 | 1.5 | -0.03 | 1.06 |
| $B_1^3$ | 140.5 | 2.8 | -0.03 | 0.97 |
| $B_1^4$ | 195.1 | 3.7 | -0.10 | 0.92 |
| $B_1^5$ | 267.5 | 4.9 | -0.06 | 0.89 |
| $B_1^6$ | 284.7 | 5.2 | -0.08 | 0.89 |

**Table S8.** Theoretical frequencies and pressure coefficients obtained from a second-order polynomial fit, $\omega=\omega_0+aP+bP^2$, for the Raman and IR-active modes in $\alpha'$-GaGeTe at 8 GPa.

| Mode | $\omega_0$ (cm$^{-1}$) | $a$ (cm$^{-1}$/GPa) | $b$ (cm$^{-1}$/GPa$^2$) |
|---|---|---|---|
| $E_g^1$ | 61.8 | 2.3 | -0.03 |
| $E_u^1$ | 84.2 | 2.1 | -0.03 |
| $A_{1g}^1$ | 116.1 | 3.2 | -0.03 |
| $A_{2u}^1$ | 129.9 | 2.0 | -0.01 |
| $E_g^2$ | 136.5 | 1.2 | - |
| $E_u^2$ | 151.5 | 3.3 | -0.03 |
| $A_{1g}^2$ | 169.4 | 0.8 | - |
| $E_g^3$ | 175.8 | 3.3 | -0.02 |
| $A_{2u}^2$ | 178.7 | 3.9 | -0.04 |
| $A_{1g}^3$ | 189.9 | 2.4 | - |

**Table S9.** Theoretical frequencies and pressure coefficients obtained from a second-order polynomial fit, $\omega=\omega_0+aP+bP^2$, for the Raman, IR, and silent modes in β'-GaGeTe at 8 GPa.

| Mode | $\omega_0$ (cm$^{-1}$) | $a$ (cm$^{-1}$/GPa) | $b$ (cm$^{-1}$/GPa$^2$) |
|---|---|---|---|
| $E_2^1$ | 31.6 | 3.1 | -0.08 |
| $E_2^2$ | 38.0 | 3.7 | -0.11 |
| $E_1^1$ | 56.1 | 6.3 | -0.18 |
| $B_1^1$ | 58.5 | 1.4 | -0.01 |
| $B_1^2$ | 69.2 | 2.7 | -0.05 |
| $E_1^2$ | 82.9 | 3.6 | -0.08 |
| $E_2^3$ | 88.6 | 4.5 | -0.11 |
| $A_1^1$ | 108.2 | 2.4 | -0.01 |
| $A_1^2$ | 129.4 | 2.2 | -0.02 |
| $B_1^3$ | 137.5 | 8.5 | -0.21 |
| $E_1^3$ | 137.7 | 0.9 | -0.01 |
| $E_2^4$ | 138.3 | -2.5 | 0.15 |
| $E_1^4$ | 150.0 | 4.0 | -0.05 |
| $E_2^5$ | 150.1 | 3.5 | -0.04 |
| $B_1^4$ | 160.0 | 5.5 | -0.12 |
| $A_1^3$ | 161.8 | 7.8 | -0.21 |
| $B_1^5$ | 167.0 | 5.0 | -0.05 |
| $E_1^5$ | 167.7 | 4.9 | -0.05 |
| $E_2^6$ | 169.9 | 5.1 | -0.07 |
| $A_1^4$ | 176.3 | 4.0 | -0.04 |
| $A_1^5$ | 183.8 | 5.9 | -0.12 |
| $B_1^6$ | 185.6 | 3.7 | -0.01 |

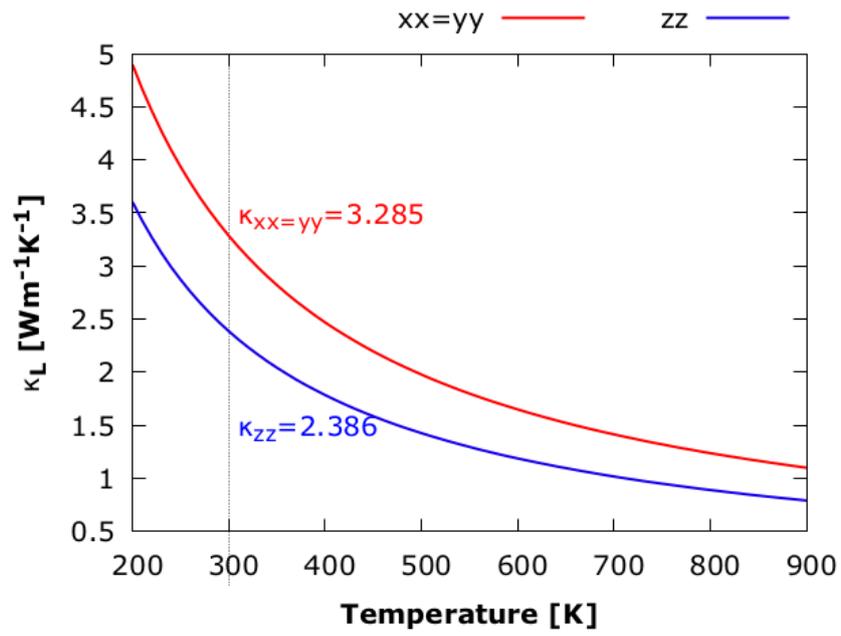

**Figure S11.** Lattice thermal conductivity ($\kappa_L$) for bulk α-GaGeTe at 5 GPa along the main directions of the hexagonal unit cell: layer plane (xx = yy) and c axis (zz).

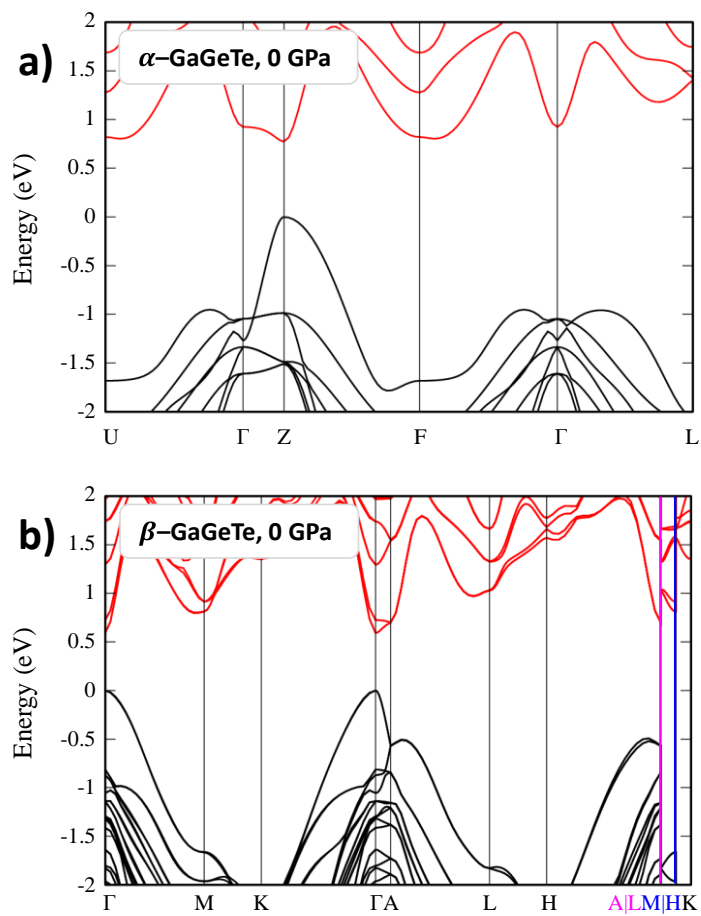

**Figure S12.** Theoretical (meta-GGA (MBJ)) electronic band structure at 0 GPa of a) $\alpha$-GaGeTe along the U-$\Gamma$-Z-F-$\Gamma$-L directions, and b) $\beta$-GaGeTe, along the the $\Gamma$-M-K-$\Gamma$-A-L-H-A-L-M-H-K directions. Given the short distance between the A, L and M, H points of the BZ of β-GaGeTe, these have been indicated as A|L (in violet) and M|H (in blue), for better readability. On the plots, the patterns corresponding to A|L and M|H are indicated in violet and in blue, respectively. The meta-GGA (MBJ were performed with the modified Becke-Johnson (MBJ) potential [2,3] that provides band gaps similar to hybrid functionals.